\documentclass[final,onefignum,onetabnum]{siamart171218}

\usepackage{lipsum}
\usepackage{amsfonts}
\usepackage{graphicx}
\usepackage{epstopdf}
\ifpdf
  \DeclareGraphicsExtensions{.eps,.pdf,.png,.jpg}
\else
  \DeclareGraphicsExtensions{.eps}
\fi

\usepackage{combelow}
\usepackage{algorithm}
\usepackage{algorithmicx}
\usepackage{algpseudocode}
\usepackage{todonotes}

\newsiamremark{remark}{Remark}
\newsiamremark{hypothesis}{Hypothesis}
\crefname{hypothesis}{Hypothesis}{Hypotheses}
\newsiamthm{claim}{Claim}

\newcommand{\obs}{{\mathrm{obs}}}
\newcommand{\sto}{{\mathrm{sto}}}

\usepackage{tikz}
\usetikzlibrary{shapes,arrows}

\headers{Multilevel Adaptive Sparse Leja Approximations}{I-G. Farca\cb{s}, J. Latz, E. Ullmann, T. Neckel, and H.-J. Bungartz}

\title{Multilevel Adaptive Sparse Leja Approximations for Bayesian Inverse Problems
\thanks{
\funding{
This work was supported by the German Research Foundation (DFG) through the TUM International Graduate School of Science and Engineering (IGSSE) within Project 10.02 BAYES, and by the EPSRC under grant number EP/R014604/1.
}}}

\author{
I-G. Farca\cb{s}\thanks{Department of Informatics, Technical University of Munich, Boltzmannstr. 3, 85748 Garching, Germany 
  (\email{farcasi@in.tum.de}, \email{neckel@in.tum.de}, \email{bungartz@in.tum.de}).}
\and 
J. Latz\thanks{Department of Mathematics, Technical University of Munich, Boltzmannstr. 3, 85748 Garching, Germany
  (\email{jonas.latz@ma.tum.de}, \email{elisabeth.ullmann@ma.tum.de}).}
\and E. Ullmann\footnotemark[3]
\and T. Neckel\footnotemark[2]
\and H.-J. Bungartz\footnotemark[2]}

\usepackage{amsopn}

\ifpdf
\hypersetup{
  pdftitle={Multilevel Adaptive Sparse Leja Approximations},
  pdfauthor={I-G. Farca\cb{s}, J. Latz, E. Ullmann, T. Neckel, and H.-J. Bungartz}
}
\fi

\begin{document}

\maketitle

% abstract
\begin{abstract}
Deterministic interpolation and quadrature methods are often unsuitable to address Bayesian inverse problems depending on computationally expensive forward mathematical models.
While interpolation may give precise posterior approximations, deterministic quadrature is usually unable to efficiently investigate an informative and thus concentrated likelihood.
This leads to a large number of required expensive evaluations of the mathematical model.
To overcome these challenges, we formulate and test a multilevel adaptive sparse Leja algorithm.
At each level, adaptive sparse grid interpolation and quadrature are used to approximate the posterior and perform all quadrature operations, respectively.
Specifically, our algorithm uses coarse discretizations of the underlying  
mathematical model to investigate the parameter space and to identify areas of high posterior probability. 
Adaptive sparse grid algorithms are then used to place points in these areas, and ignore other areas of small posterior probability.
The points are weighted Leja points. 
As the model discretization is coarse, the construction of the sparse grid is computationally efficient.
On this sparse grid, the posterior measure can be approximated accurately with few expensive, fine model discretizations. 
The efficiency of the algorithm can be enhanced further by exploiting more than two discretization levels.
We apply the proposed multilevel adaptive sparse Leja algorithm in numerical experiments involving elliptic inverse problems in 2D and 3D space, in which we compare it with Markov chain Monte Carlo sampling and a standard multilevel approximation.
\end{abstract}
% REQUIRED

% REQUIRED

% introduction
\section{Introduction} \label{Sec_Intro}
Mathematical models in science and engineering often require input parameters which cannot be observed directly, yet these parameters are required for predictions based on the model.
A standard procedure is to estimate the inputs from indirect observations, which is known as an \textit{inverse problem}.
In contrast, the corresponding \textit{forward problem} maps from the parameter to the observation space.

In many applications, for instance the geosciences or medical sciences, the observations are noisy and their number is insufficient to identify a unique associated parameter value.
The Bayesian approach to inverse problems \cite{Kaipio2005, AMS10, Tarantola05} provides a consistent mechanism to combine noisy or incomplete data with prior knowledge, and to quantify the uncertainty in the parameter estimate.
The prior knowledge is incorporated into a probability distribution over the parameter space; this is termed \textit{prior (measure)} $\mu_0$.
The Bayesian solution to the inverse problem is then the \textit{posterior (measure)} $\mu^{\boldsymbol y}$ arising from conditioning the prior $\mu_0$ on the observations.
Unfortunately, the posterior is often intractable in the sense that it does not admit closed form analytic expressions.
Hence approximations have to be used in practice.

\textit{Sampling-based} posterior approximations such as Markov chain Monte Carlo (MCMC) \cite{HandbookMCMC} or Sequentical Monte Carlo (SMC) \cite{DelMoral2006} do not rely on the smoothness of the parameter-to-observation map, and can be conducted in high-dimensional parameter spaces. 
The drawback is that without exploiting smoothness or low-dimensional structure in the parameter space often a prohibitive number of samples are required to obtain the desired accuracy. 
Since each sample entails the evaluation of the forward map, the total cost of the Bayesian inversion becomes prohibitive if the forward model is specified by a partial differential equation (PDE), and is thus computationally expensive.
In recent years, many works suggested computationally feasible, yet accurate approximations of the posterior.
In this work we focus on \textit{sampling-free} approximations for computationally expensive problems for parameter spaces of small to moderate dimension.
We assume that the prior and posterior have a probability density function with respect to (w.r.t.) the Lebesgue measure and approximate the posterior density using sparse grid interpolation and sparse grid quadrature. 

Sampling-free approaches often involve \textit{surrogates} for the forward response operator to decrease the computational cost.
Typical surrogates are based on generalized polynomial chaos \cite{Eigel2018, LiMarzouk:2014, MarzoukXiu:2009, YanZhou:2018}, sparse grids \cite{ChenSchwab:2015, ChenVillaGhattas:2017, MZ09, SchillingsSchwab:2013, SchillingsSchwab:2014, SchwabStuart:2012}, Gaussian process regression \cite{Kennedy01, StuartTeckentrup:2018}, model reduction \cite{Frangos11, LEU18, Lieberman10}, and combinations, e.g., sparse grids and reduced bases \cite{ChenSchwab:2015, ChenSchwab:2016a}. 
To obtain an accurate (and convergent) approximation, these surrogates require certain types of smoothness of the response surface or likelihood function w.r.t.~ the input parameters.
The smoothness assumptions can be weakened by using piecewise polynomial approximations together with Voronoi tesselations of the parameter space \cite{Mattis2018}.
Of course, surrogates can also be used to accelerate sampling-based approximations such as MCMC, see e.g., \cite{LiMarzouk:2014,PM18}.
We remark that Quasi-Monte Carlo \cite{Di17, SST17} is in principle a sampling-free method which does not rely on surrogates, however, it requires again a smooth approximand, and is often used together with randomization.

Theoretical analysis shows that if the surrogate converges to the forward model at a specific rate w.r.t. the prior weighted $L^2$-norm, then the approximate posterior converges to the exact posterior with at least the same rate \cite{MarzoukXiu:2009, AMS10}.
This result has been improved recently in \cite{YanZhang:2017} where it was showed that the convergence rate of the posterior approximation is at least twice as large as the convergence rate of the surrogate, for general priors.
However, constructing an accurate surrogate over the entire support of the prior might not be feasible and is in fact often unnecessary.
Indeed, in inference problems where the data is informative, the posterior differs significantly from the prior, and is supported only in a small part of the prior support.
This suggests to adapt and localize the surrogate construction to the support of the posterior.
We adopt this approach in our work and construct multilevel, adaptive surrogates with localized support using adaptive sparse grid approximations.

The idea of posterior-focused surrogates is not new, it has however received little attention to date in the literature.
Li and Marzouk \cite{LiMarzouk:2014} borrow ideas from statistics and construct an efficient polynomial chaos surrogate associated with a density that minimizes the cross entropy between the posterior and a family of multivariate normal distributions.
Jiang and Ou \cite{JiangOu:2018} suggest a two-stage surrogate based on generalized multiscale finite elements and least-squares stochastic collocation.
Yan and Zhou \cite{YanZhou:2018} propose a multifidelity polynomial chaos surrogate which combines a large number of inexpensive low-fidelity model evaluations with a small number of expensive high-fidelity model evaluations, following the idea of multifidelity approximations \cite{Peherstorfer:2018}.

One challenge of posterior-focused surrogates is the need to handle arbitrary densities which can deviate significantly from the prior which is usually a classical density such as uniform or Gaussian.
We address this by constructing adaptive sparse grid approximations based on \emph{weighted (L)-Leja sequences} (see, e.g., \cite{GO16, NJ14}).
%In particular, weighted (L)-Leja sequences are used in adaptive sparse grid interpolation and quadrature to respectively approximate the posterior and perform all quadrature operations.
Note that sparse grid approximations with Leja points have been devised for forward uncertainty propagation in \cite{Fa18b, Fa18, Geo18, NJ14}.
In \cite{Fa18b} the adaptive construction of the points is guided by sensitivity scores, and the strategy is applied in plasma microturbulence analysis.
In \cite{Geo18} the Leja points are constructed with the help of an adjoint-based error indicator.
The use of Leja points to approximate posterior densities is a novel contribution to the literature. 
Leja points offer further computational advantages since they are \textit{nested} and thus allow to reuse (expensive) model evaluations.

We address the possible high-dimensional parameter space in Bayesian inversion by the use of multilevel approximations.
At each level, \textit{dimension-adaptive} sparse grids \cite{CM13, GG03, NJ14} are employed, either in standard form, or using directional variances to better exploit anisotropies in the parameter space.
In particular, at the first level, our  algorithm uses a coarse discretization of the given model to investigate the parameter space and to identify areas of high posterior probability. 
Adaptive algorithms are then used to place weighted (L)-Leja points in these areas, and ignore other areas of small posterior probability.
Starting with the second level, we sequentially update the prior such that the previous sparse grid approximation of the posterior is reused. 
In this way, the current posterior measure can be approximated accurately with few expensive, finer model discretizations.
We point out that sparse grid approximations are based on point sequences in one dimension. 
However, starting with the second level the posterior densities are in general not separable and hence we cannot rely on a simple tensorization of univariate Leja points. 
Instead we construct Leja points w.r.t.~ a \textit{Gaussian approximation} of the posterior, which is separable, and we then correct the bias introduced by this approximation in quadrature computations.

%Alternative approaches include transport maps \cite{Ma16} and copulas.
%Moreover, it would be possible to carry out the Leja construction in the multi-dimensional parameter space. 
%However, as the dimension increases the optimization problem which is required for the Leja construction becomes more expensive and eventually infeasible.

%
%
%\subsection{Challenges}
%\begin{itemize}
%	\item
%	how to construct accurate surrogates of likelihood?
%	\item
%	adaptive strategies for Bayesian inversion, problem is the concentrated posterior, adaptive algorithm might stop early
%	\item
%	nested point sets for efficieny; use Leja points
%	\item
%	need sparse grids for arbitrary densities; again Leja, density adaptivity
%	\item
%	non-separable intermediate or posterior densities; address this by Gaussian approximation; alternatives are transport maps, copulas, Leja points in d-dimensional space (Jakeman, expensive optimization problem)
%	\item
%	high-dimensional parameter space; address this by dimension adaptivity (standard and directional variance; minor novel contribution)
%	\item
%	expensive likelihood evaluation; address this by multilevel Leja construction (major contribution of this work)
%\end{itemize}

The remainder of this paper is structured as follows.
In \cref{sec:background} we provide the necessary background information.
In particular, in \cref{subsec:bayesian} we formulate the Bayesian inverse problem and discuss the computation of posterior expectations by importance sampling.
\Cref{subsec:ml_overview} reviews multilevel approaches, and \cref{subsec:sparse_grids} discusses generalized sparse grids.
\Cref{sec:ml} contains the major contribution of our work, the multilevel adaptive sparse Leja approximation to the posterior density in a Bayesian inverse problem.
This method is independent of the specific implementation of the adaptive sparse grid approximations. 
Details on the used dimension-adaptive approximations are given in \cref{sec:dim_adapt}.
In \cref{sec:results}, we present numerical results, comparing our multilevel algorithm with sampling methods and the classical multilevel approach based on telescoping sums.
Finally, \cref{sec:conc} offers concluding remarks.

% Distinguish sampling-free and sampling-based
% Sampling-based: MCMC, SMC
% Sampling-free: tensor stuff by Eigel et al., QMC by Gantner et al., PC, Voronoi, Gaussian processes (Aretha)

% Distinguish exact model evaluations vs surrogates, different types of surrogates e.g. based on sparse grids, PC, construction with respect to prior measure
% Distinguish single-level and multilevel approaches, 
% Distinguish adaptive (mattis, Li&Marzouk) vs non-adaptive

% Discuss surrogate convergence w.r.t. prior (Yan&Zhang), discuss concentration of posterior (Marzouk 2009)
% Discuss arbitrary densities, Leja points

% Discuss output quantities of interest, our algorithm gives polynomial density approximation, works also for non-PDE-based models; require only a hierarchy of models (multifidelity)

% Discuss challenges of surrogate construction in Bayesian inversion
% Our method has similarities to importance sampling / a change of measure

% In conclusions: multimodal posteriors, goal-oriented surrogate construction, 
% convergence analysis of multilevel surrogates, Yan&Zhang show that we need a more accurate surrogate if the noise variance is small.

% Bayesian inversion
\section{Background} \label{sec:background}
\subsection{Bayesian inversion}\label{subsec:bayesian}

To begin we formulate the Bayesian inverse problem.
% and state the assumptions on the parameter space and prior required by our algorithm.
Let $\boldsymbol{X} \subset \mathbb{R}^{N_\sto}$ denote the parameter space. 
In addition, let  $\boldsymbol{Y} = \mathbb{R}^{N_\obs}$ be a separable Banach space that denotes the data space. 
$N_\sto$ is the dimension of the data space, and $N_\obs$ is the number of observations. 
Notably, the parameter and data space are \textit{finite-dimensional}.
This allows us to work with densities w.r.t. the Lebesgue measure.
The underlying mathematical model is formalized by a function $\mathcal{G}: \boldsymbol{X} \rightarrow \boldsymbol{Y}$, which maps from the parameter space to the data space.
Noisy observations $\boldsymbol{y} \in \boldsymbol{Y}$ are obtained.
To model the noise we assume that $\boldsymbol{y}$ is a realisation of the random variable $\mathcal{G}(\boldsymbol{\theta}^{\mathrm{true}})+\boldsymbol{\eta}$ where $\boldsymbol{\eta} \sim \mathrm{N}(0, \Gamma)$ is non-degenerate Gaussian noise, and $\boldsymbol{\theta}^{\mathrm{true}} \in \boldsymbol{X}$ is the true parameter.
In an inverse problem we wish to identify the parameter $\boldsymbol{\theta}^{\mathrm{true}}$, i.e., solve the equation
\begin{equation}\label{eq:Inverse_Problem}
\mathcal{G}(\boldsymbol{\theta})+\boldsymbol{\eta} = \boldsymbol{y}
\end{equation}
for $\boldsymbol{\theta}$.
This problem is typically ill-posed in the sense of Hadamard \cite{JH1902}, due to noise, and since the low-dimensional data space is often not sufficiently rich to allow the identification of a unique parameter in the high-dimensional space $\boldsymbol{X}$. 
The ill-posedness can be cured by reformulating \cref{eq:Inverse_Problem} as a Bayesian inverse problem.
Next, we introduce our notation and briefly discuss Bayesian inversion, and refer to \cite{AMS10} for more details.

We assume that $\boldsymbol{\theta}$ is an $\boldsymbol{X}$-valued random variable that is distributed according to a \emph{prior measure} $\mu_0$ with Lebesgue density $\pi_0$ on the parameter space $\boldsymbol{X}$.
Moreover, we assume that  $\boldsymbol{\theta}$ is stochastically independent of the noise $\boldsymbol{\eta}$.
The density $\pi_0$ reflects our knowledge about $\boldsymbol{\theta}$ before we make an observation  $\boldsymbol{y}$. 
The information provided by $\boldsymbol{y}$ is modelled by the (data) likelihood.
Since the noise $\boldsymbol{\eta}$ is Gaussian by assumption, the likelihood is given by
\begin{align}
L(\boldsymbol{\theta}|\boldsymbol{y}) &:= \exp\left(-\Phi(\boldsymbol{\theta};\boldsymbol{y})\right),  \nonumber \\
\Phi(\boldsymbol{\theta};\boldsymbol{y}) &:= \frac{1}{2}\|\Gamma^{-1/2}(y-\mathcal{G}(\boldsymbol{\theta}))\|^2  \label{eq:potential}.
\end{align} 
The function $\Phi$ is called potential or negative log-likelihood. 
%We omit the dependence on the data $\boldsymbol{y}$ and write $\Phi := \Phi(\cdot, \boldsymbol{y})$.
The solution of the Bayesian inverse problem is the \emph{posterior (measure)}  $\mu^{\boldsymbol{y}}$, i.e., the conditional measure of $\boldsymbol{\theta}$ given that the event $\{\mathcal{G}(\boldsymbol{\theta}) + \boldsymbol{\eta} = \boldsymbol{y}\}$ occurred.
The posterior measure $\mu^{\boldsymbol{y}}$ has also a density $\pi^{\boldsymbol{y}}$ which can be computed using Bayes's formula: 
\begin{align}
\pi^{\boldsymbol{y}}(\boldsymbol{\theta}) 
&= \frac{L(\boldsymbol{\theta}|\boldsymbol{y})\pi_0(\boldsymbol{\theta})}{Z(\boldsymbol{y})}, \qquad \boldsymbol{\theta} \in X,\ \boldsymbol{y} \in Y, \label{eq:posterior} \\
Z(\boldsymbol{y}) 
&= \int_{\boldsymbol{X}} L(\boldsymbol{\theta}|\boldsymbol{y})\pi_0(\boldsymbol{\theta}) \mathrm{d}\boldsymbol{\theta} \nonumber,
\end{align}
provided that $0 < Z(\boldsymbol{y}) < \infty$.
In the given setting (non-degenerate Gaussian additive noise, finite dimensional data space), one can show that $Z(\boldsymbol{y})$ is always finite and bounded away from $0$. 
This implies existence of the posterior measures, see \cite{Latz2019}.
The work \cite{Latz2019} also establishes that Bayesian inverse problems of this type are always \emph{well-posed}.

Finally, let $g: \boldsymbol{X} \rightarrow \mathbb{R}$ denote a quantity of interest (QoI) depending on the parameter $\boldsymbol{\theta}$.
Since $\boldsymbol{\theta}$ is a random variable, one is typically interested in the forward propagation of uncertainties through the action of ${g}$. 
For instance, we wish to evaluate integrals of $g$ w.r.t. the posterior measure:
\begin{equation} \label{eq:qoi}
\mathbb{E}_{\mu^{\boldsymbol{y}}}[g] 
= 
\int_{\boldsymbol X} g(\boldsymbol{\theta}) \pi^{\boldsymbol{y}}(\boldsymbol{\theta}) \mathrm{d}\boldsymbol{\theta}.
\end{equation}
In practice, we approximate integrals of this type via numerical quadrature. 
Note that we can write the expected value in \cref{eq:qoi} in terms of a ratio of two expected values w.r.t. the prior measure:
 \begin{equation} \label{eq:ImportSampIdent}
 \mathbb{E}_{\mu^{\boldsymbol{y}}}[g] 
 = 
% \int g(\boldsymbol{\theta}) \pi^{\boldsymbol{y}}(\boldsymbol{\theta}) \mathrm{d}\boldsymbol{\theta} 
% = 
 \frac{1}{Z(\boldsymbol{y})} \int g(\boldsymbol{\theta})L(\boldsymbol{\theta}|\boldsymbol{y})\pi_0(\boldsymbol{\theta}) \mathrm{d}\boldsymbol{\theta} 
 = 
 \frac{\mathbb{E}_{\mu_0}[g(\cdot) L(\cdot|\boldsymbol{y})]}{\mathbb{E}_{\mu_0}[L(\cdot|\boldsymbol{y})]}.
  \end{equation}  
Note that the prior is typically much more accessible compared to the posterior, via i.i.d. sampling or a closed form probability density function (pdf).  
If the two expected values in \cref{eq:ImportSampIdent} are approximated with samples from $\mu_0$, we refer to this method by \emph{importance sampling}. %see e.g. \cite{APSS17}.
If the expected values are approximated with other  numerical quadrature rules, e.g., Quasi-Monte Carlo \cite{SST17} or sparse grid quadrature \cite{SchillingsSchwab:2013,SchwabStuart:2012}, we refer to any of these methods as \emph{importance-sampling-based methods}.

%\subsection{Multilevel and multifidelity}\label{subsec:ml_overview}
\subsection{Multilevel approximation of the quantity of interest}\label{subsec:ml_overview}

The approximation of the expected value $\mathbb{E}_{\mu^{\boldsymbol{y}}}[g]$ involves two sources of error: $(i)$ the quadrature error associated with the approximation of the measure $\mu^{\boldsymbol{y}}$, and $(ii)$ the discretization error associated with the approximation of $g$.
The quadrature error is typically controlled by the number of particles or grid points in the parameter domain.
The discretization error in $g$ is often controlled by the mesh size in the physical domain.  
In our setting, the evaluation of $g$ typically involves a computationally expensive model, e.g., a partial differential equation (PDE).
If we wish to construct accurate approximations to $\mathbb{E}_{\mu^{\boldsymbol{y}}}[g]$, then $g$ has to be discretized with a high resolution on a fine grid in physical space, and $g$ has to be evaluated for a large number of parameter values.
For these reasons the approximation of $\mathbb{E}_{\mu^{\boldsymbol{y}}}[g]$ is computationally demanding.
 
\emph{Multilevel methods} provide a framework to approximate high-resolution problems efficiently by combining evaluations from fine and coarse grid approximations.
Specifically, approximations based on quadrature and physical domain discretization grids of complementary resolution are combined such that most computations are done on coarse grids while fine grid approximations are evaluated only a limited number of times.
In this way, the overall computational cost is reduced, while the accuracy is preserved.
%\emph{Multifidelity} refers to the use of different surrogates or physical models for the forward problem.
%In the following, we do not distinguish between multilevel and {multifidelity} methods.
A large number of multilevel approaches for Bayesian inverse problems proceed by using a telescoping sum based on the linearity of the expectation operator, see, e.g., \cite{DKST15}.
Alternatively, it is possible to construct a multilevel approximation without relying on such a telescoping sum.
Examples are the multifidelity preconditioned MCMC method in \cite{PM18}, Multilevel Sequential${}^2$ Monte Carlo \cite{LPU18}, and our multilevel sparse Leja approximation presented in \cref{sec:ml}.

\subsection{Approximation with generalized sparse grids} \label{subsec:sparse_grids}
We aim to approximate posterior density functions with sparse grid interpolation and quadrature at each level in our multilevel approach.
To this end, we employ \emph{generalized, adaptive Smolyak approximations} \cite{Sm63}.
Smolyak's algorithm, also known as the combination scheme \cite{GSZ92}, is a strategy to construct multivariate sparse grid approximations by weakening the assumed coupling between the input dimensions (see, e.g., \cite{CM13, Fa18b}). 
We briefly summarize Smolyak's algorithm. 
For a more detailed overview of this approach and its application in scientific computing, see \cite{BG04,TW18} and the references therein. 

Let $f^i \colon X_i \subset \mathbb{R} \rightarrow \mathbb{R}$ denote univariate functions, where $i = 1, 2, \ldots, N_\sto$.
In addition, let $f^{\boldsymbol{N_\sto}}: \boldsymbol{X} \rightarrow \mathbb{R}$ denote a multivariate function with a scalar output.
For example, $f^{\boldsymbol{N_\sto}}$ could be the potential function, $\Phi(\boldsymbol{\theta};\boldsymbol{y})$, or the QoI, $g(\boldsymbol{\theta})$.
Let $\mathcal{U}^{i}[f^i]$ for $i = 1, 2, \ldots, N_\sto$ denote either univariate interpolation or integration operators defined w.r.t. a weight function $w: X_i \rightarrow \mathbb{R}_+$, which in our context is the $i$th component of the prior density.
Further, consider approximations $\mathcal{U}^{i}_k[f^i]$ that converge as $k \rightarrow \infty$, where $k$ is typically referred to as \emph{level}.
Starting from one-dimensional \emph{difference} or \emph{hierarchical surplus operators},
\begin{equation}\label{eq:univar_delta_operator}
\Delta^{i}_k[f^i] := \mathcal{U}^{i}_k[f^i] - \mathcal{U}^{i}_{k - 1}[f^i], \quad i = 1, 2, \ldots, N_\sto,
\end{equation}
with the convention $\Delta^{i}_1[f^i] := \mathcal{U}^{i}_1[f^i]$, Smolyak's approximation formula reads
\begin{equation} \label{eq:smolyak_formula}
\boldsymbol{\mathcal{U}}_{\mathcal{K}}[f^{\boldsymbol{N_\sto}}] = 
\sum_{\boldsymbol{k} \in \mathcal{K}} (\Delta^{1}_{k_1} \otimes \Delta^{2}_{k_2} \otimes \ldots \otimes  \Delta^{N_\sto}_{k_{N_\sto}})[f^{\boldsymbol{N_\sto}}] =
 \sum_{\boldsymbol{k} \in \mathcal{K}} \boldsymbol{\Delta}_{\boldsymbol{k}}[f^{\boldsymbol{N_\sto}}],
\end{equation}
where $\boldsymbol{k} := (k_1, k_2, \ldots, k_{N_\sto}) \in \mathbb{N}^{N_\sto}$ is a \emph{multiindex} and $\mathcal{K}$ is a \emph{finite set of multiindices}.
Note that by construction, \cref{eq:smolyak_formula} requires the underlying multivariate space, $\boldsymbol{X}$, as well as the corresponding weight function to be separable.
When the weight function is a density the stochastic parameters need to be independent.

Since \cref{eq:smolyak_formula} is written in terms of tensorizations of univariate difference operators \cref{eq:univar_delta_operator} the set $\mathcal{K}$ must be constructed such that the summation in \cref{eq:smolyak_formula} telescopes correctly. 
Suitable sets of multiindices $\mathcal{K}$ are called \emph{admissible} (or \emph{downward closed}; see \cite{GG03}).
In particular, for an admissible set $\mathcal{K}$ it holds that $\boldsymbol{k} \in \mathcal{K} \Rightarrow \boldsymbol{k} - \boldsymbol{e}_i \in \mathcal{K}$ for $i = 1, 2, \ldots, {N_\sto}$, where $\boldsymbol{e}_i$ denotes the $i$th unit vector in $\mathbb{R}^{N_\sto}$. 
%For computations, \cref{eq:smolyak_formula} is typically rewritten as 
%\begin{equation*}
%\boldsymbol{\mathcal{U}}_{\mathcal{K}}[f^{\boldsymbol{N_\sto}}] = \sum_{\boldsymbol{k} \in \mathcal{K}} c_{\boldsymbol{k}} \boldsymbol{\mathcal{U}}_{\boldsymbol{k}}[f^{\boldsymbol{N_\sto}}],
%\end{equation*}
%where $c_{\boldsymbol{k}} = \sum_{\boldsymbol{z} \in \{0, 1\}^{N_\sto}} (-1)^{|\boldsymbol{z}|_1} \chi_{\mathcal{K}}(\boldsymbol{k} + \boldsymbol{z})$ and $\chi_{\mathcal{K}}$ is the indicator function on $\mathcal{K}$, i.e., 
%$\chi_{\mathcal{K}}(\boldsymbol{k}) = 1$ if $\boldsymbol{k} \in \mathcal{K}$ and $\chi_{\mathcal{K}}(\boldsymbol{k}) = 0$ otherwise. 

%We summarize the interpolation and quadrature operators in \cref{subsec:operators}.
To construct the approximations $\mathcal{U}^{i}_k[f^i]$, we employ \emph{weighted (L)-Leja sequences} (see, e.g., \cite{GO16, NJ14}).
Given the weight function $w: X_i \rightarrow \mathbb{R}_+$, weighted (L)-Leja sequences are constructed recursively as follows:
\begin{equation}\label{eq:weighted_leja_points}
\begin{split}
& \theta_1 = \underset{\theta \in X_i}{\mathrm{argmax}}{\ w(\theta)}, \\
\ & \theta_n = \underset{\theta \in X_i}{\mathrm{argmax}} \ w(\theta)\prod_{m=1}^{n-1}(\theta - \theta_m), \quad n = 2, 3, \ldots
\end{split}
\end{equation}
When $w$ is the standard uniform density with support $[0, 1]$, we choose $\theta_1 = 0.5$.
Note that the above point sequence is in general not uniquely defined, because \cref{eq:weighted_leja_points} might have multiple maximizers.
In that case we simply pick one of the maximizers.
Weighted (L)-Leja sequences allow the construction of sparse grid approximations for arbitrary probability densities.
Moreover, they lead to accurate approximations with low cardinality (see, e.g., \cite{NJ14}).
To fully define \cref{eq:smolyak_formula}, we need to specify the multiindex set $\mathcal{K}$ as well.
We construct $\mathcal{K}$ adaptively based on the \emph{dimension-adaptive} algorithm of \cite{GG03, He03}, which we outline in \cref{sec:dim_adapt}.

\subsection{Assumptions}
As discussed in \cref{subsec:sparse_grids}, sparse grid approximations require a tensor domain and a tensorized prior measure. 
Hence, we assume:
\medskip
\begin{itemize}
\item[\bf A1.]
The parameter space is a tensor product space, i.e., $$\boldsymbol{X}= \bigotimes_{i=1}^{N_\sto}X_i,$$ where $X_i \subset \mathbb{R}$ for $i=1,\dots,N_\sto$. \medskip
\item[\bf A2.]
The prior density $\pi_0$ is separable, i.e.,
\begin{equation}\label{re:remark_indep}
\pi_0(\boldsymbol{\theta}) 
=  
\prod_{i=1}^{N_\sto}\pi_{0,i}(\theta_i),
\end{equation}
where $\pi_{0,i} \colon X_i \rightarrow \mathbb{R}$, $i=1,\dots,N_\sto$. 
\end{itemize}\medskip
Assumption A1 can always be satisfied by embedding a non-tensorized parameter space into a hyperrectangle of suitable dimension.
Assumption A2 is fulfilled if the components of $\boldsymbol \theta$ are stochastically independent under the prior measure.
If Assumption A2 is not satisfied, the Gaussian approximation of $\pi_0$ is needed (see \cref{sec:ml}).

% sparse grids
\section{Multilevel sparse Leja algorithm}\label{sec:ml}
This section contains the major contribution of our work.
Our goal is to address the challenges of Bayesian inversion in computationally expensive problems. 
To this end, we formulate a deterministic, multilevel, sampling-free methodology based on sparse grids in which we sequentially update the prior information as the level in the multilevel hierarchy increases.

Most computations in Bayesian inversion involve evaluations of the forward operator, $\mathcal{G}(\boldsymbol{\theta})$, which in this paper is assumed to be computationally expensive. 
When a large number of such evaluations is needed, the corresponding computational cost is prohibitive.
To reduce the cost, we employ multilevel approximations.
At each level, we construct sparse grid surrogates of the potential function $\Phi(\boldsymbol{\theta};\boldsymbol{y})$.
Our motivation is two-fold.
First, evaluating $\Phi(\boldsymbol{\theta};\boldsymbol{y})$ means evaluating the forward model, $\mathcal{G}(\boldsymbol{\theta})$ (see \cref{eq:potential}), hence the computationally expensive part.
Second, even if $\mathcal{G}(\boldsymbol{\theta})$ is vector valued, $\Phi(\boldsymbol{\theta};\boldsymbol{y})$ is a scalar.
Sparse grid approximations can be constructed for vector-valued functions, however, a separate approximation is usually needed for each output component, which can be infeasibly expensive. 
After the surrogate is obtained, all other single level operations, which typically involve integration, employ this surrogate, making them computationally cheap.
In the following, we use the superscripts $\emph{in}$ and $\emph{qu}$ to specifically refer to interpolation and quadrature, respectively, whereas superscript $\emph{op}$ is used to refer to either of the two operations.

In this paper, the surrogates of the potential function are constructed via adaptive sparse grid interpolation, whereas all integration operations are performed using adaptive sparse grid quadrature.
The specific implementations do not influence the formulation of our multilevel approach. 
Thus we assume we have two adaptive strategies, \texttt{AdaptSGInterp}($tol^{\mathrm{in}}$, $K_{\mathrm{max}}^{\mathrm{in}}$, $g$, $\pi$) and \texttt{AdaptSGQuad}($tol^{\mathrm{qu}}$, $K_{\mathrm{max}}^{\mathrm{qu}}$, $g$, $\pi$), which depend on a \emph{tolerance}, $tol^{\mathrm{op}}$.
The other inputs are a maximum reachable sparse grid level, $K_{\mathrm{max}}$, the target function, $g$, and the density function w.r.t.which the approximation is performed, $\pi$.
Note that specific implementations might have additional input arguments, however the four inputs considered here are sufficient to illustrate these algorithms.
The adaptive strategies are summarized in \cref{sec:dim_adapt}.

\subsection{General setup}\label{subsec:our_approach}
Let $J > 1$, $J \in \mathbb{N}$, denote the number of \emph{levels} in our multilevel formulation.
Further, let $A \in \{\mathcal{G}, \Phi, L, Z, \pi^{\boldsymbol{y}}\}$ denote a generic quantity depending on both physical and stochastic parameters.
Let $j = 1, 2, \ldots, J$.
By $h_j$ we characterize the discretization of the physical domain of the forward response operator $\mathcal{G}(\boldsymbol{\theta})$, where $h_1$ is the coarsest and $h_J$ the finest discretization level.
Hence, by $A_{j}$ we denote the semi-discrete approximation of $A$ depending on $h_j$, whereas $A_{\delta j}$ denotes either $A_{j} - A_{j - 1}$ or $A_{j}/A_{j - 1}$.
In addition, $tol^{\mathrm{op}}_j$ denotes the tolerance employed in the adaptive sparse grid approximations of $A_j$ such that \[tol^{\mathrm{op}}_1 \leq tol^{\mathrm{op}}_2 \leq \ldots \leq tol^{\mathrm{op}}_J.\]
Thus by $A_{j, s}$ we denote the sparse grid approximation of $A_j$ depending on $tol^{\mathrm{op}}_s$.

In our multilevel formulation, we determine $A_{j, J - j + 1}$ for all $j = 1, 2, \ldots, J$.
To simplify the notation we use the subscript $\ell(j)$ to refer to $(h_j, tol^{\mathrm{op}}_{J - j + 1})$ and the subscript $\ell(\delta j)$ to denote approximations $A_{\ell(\delta j)} \approx A_{\delta j}$.
Hence we refer to the \emph{level $j$} in our multilevel approach by $\ell(j)$ or $\ell(\delta j)$.
Note that levels are used to characterize both sparse grid and multilevel formulations. 
To avoid confusion, we will explicitly specify what is meant by level in each context.

\subsection{Level $\ell(1)$} \label{subsubsec:level_1}
At $\ell(1)$ we compute the approximation $\Phi_{\ell(1)}(\boldsymbol{\theta};\boldsymbol{y}) \approx \Phi_1(\boldsymbol{\theta};\boldsymbol{y})$ using adaptive sparse grid interpolation w.r.t. the prior $\pi_0$.
This is possible since $\pi_0$ has a product structure by Assumption A2, see \cref{re:remark_indep}. 
The potential's approximation $\Phi_{\ell(1)}(\boldsymbol{\theta};\boldsymbol{y})$ is used for $L_1(\boldsymbol{\theta}|\boldsymbol{y}) \approx L_{\ell(1)}(\boldsymbol{\theta}|\boldsymbol{y}) := \exp{(-\Phi_{\ell(1)}(\boldsymbol{\theta};\boldsymbol{y}))}$, the level one likelihood surrogate.
Afterwards, we employ $L_{\ell(1)}(\boldsymbol{\theta}|\boldsymbol{y})$ to compute the evidence $Z_{\ell(1)}(\boldsymbol{y}) \approx Z_1(\boldsymbol{y})$ via adaptive sparse grid quadrature w.r.t. the prior density $\pi_0$.
Having computed $L_{\ell(1)}(\boldsymbol{\theta}|\boldsymbol{y})$ and $Z_{\ell(1)}(\boldsymbol{y})$ we apply formula \cref{eq:posterior} and obtain the posterior at level $\ell(1)$, $\pi_{\ell(1)}^{\boldsymbol{y}}(\boldsymbol{\theta})$.
In addition, we also compute the posterior expectation,
$\boldsymbol{m}_{\ell(1)} \in \mathbb{R}^{N_\sto}$, where for $i = 1, \ldots, N_\sto$,
\begin{equation*}
\boldsymbol{m}_{\ell(1)}^i 
:= 
 Z_{\ell(1)}^{-1}(\boldsymbol{y})
\int_{\boldsymbol{X}} \theta_i L_{\ell(1)}(\boldsymbol{\theta}|\boldsymbol{y}) \pi_0(\boldsymbol{\theta}) \mathrm{d} \boldsymbol{\theta}
\end{equation*} 
and covariance matrix, $\boldsymbol{C}_{\ell(1)} \in \mathbb{R}^{N_\sto \times N_\sto}$, where for $n, p = 1, \ldots, N_\sto$,
\begin{equation*}
\boldsymbol{C}_{\ell(1)}^{np} 
:= 
Z_{\ell(1)}^{-1}(\boldsymbol{y})
\int_{\boldsymbol{X}} \theta_n \theta_p L_{\ell(1)}(\boldsymbol{\theta}|\boldsymbol{y}) \pi_0(\boldsymbol{\theta}) \mathrm{d} \boldsymbol{\theta} - \boldsymbol{m}_{\ell(1)}^n \boldsymbol{m}_{\ell(1)}^p,
\end{equation*}
which we need to construct the \emph{Gaussian approximation} $\widehat{\pi}_{\ell(1)}^{\boldsymbol{y}}(\boldsymbol{\theta}) := \mathrm{N}(\boldsymbol{m}_{\ell(1)}, \boldsymbol{C}_{\ell(1)})$ of $\pi_{\ell(1)}^{\boldsymbol{y}}(\boldsymbol{\theta})$ at the second level $\ell(2)$ (see \cref{subsubsec:gaussian_approx}).

These steps are summarized in \cref{algo:our_approach_step_1}.
The first three inputs are the spatial discretization parameter $h_1$ and the tolerances for sparse grid interpolation and quadrature, $tol^{\mathrm{in}}_1$ and $tol^{\mathrm{qu}}_1$. 
Moreover, $\boldsymbol{K}_{\mathrm{max}} := (K_{\mathrm{max}}^{\mathrm{in}}, K_{\mathrm{max}}^{\mathrm{qu}})$ comprises the maximum attainable sparse grid levels for the two adaptive operations.
The last two inputs are the potential function, $\Phi(\boldsymbol{\theta};\boldsymbol{y})$, and the prior density, $\pi_0(\boldsymbol{\theta})$.
\begin{algorithm}
\caption{Level One Adaptive Sparse Leja Algorithm for Bayesian Inversion}\label{algo:our_approach_step_1}
\begin{algorithmic}[1]
\Procedure{Level1SparseLeja}{$h_1, tol^{\mathrm{in}}_1, tol^{\mathrm{qu}}_1, \boldsymbol{K}_{\mathrm{max}}, \Phi(\boldsymbol{\theta};\boldsymbol{y}), \pi_0(\boldsymbol{\theta})$}
\State \label{sl_begin} {Compute the potential's surrogate via adaptive sparse grid interpolation} 
\begin{equation*}
\Phi_{\ell(1)}(\boldsymbol{\theta};\boldsymbol{y}) = \mathrm{\texttt{AdaptSGInterp}}(tol_{1}^{\mathrm{in}}, K_{\mathrm{max}}^{\mathrm{in}}, \Phi_{1}, \pi_0)
\end{equation*}
\State Construct the likelihood surrogate $L_{\ell(1)}(\boldsymbol{\theta}|\boldsymbol{y}) := \exp{(-\Phi_{\ell(1)}(\boldsymbol{\theta};\boldsymbol{y}))}$
\State Compute the evidence via adaptive sparse grid quadrature
\begin{equation*}
Z_{\ell(1)}(\boldsymbol{y}) = \mathrm{\texttt{AdaptSGQuad}}(tol_{1}^{\mathrm{qu}}, K_{\mathrm{max}}^{\mathrm{qu}}, L_{\ell(1)}, \pi_0)
\end{equation*}
\State Compute the posterior \label{sl_end}
\begin{equation*}
\pi_{\ell(1)}^{\boldsymbol{y}}(\boldsymbol{\theta}) := \frac{\pi_0(\boldsymbol{\theta})L_{\ell(1)}(\boldsymbol{\theta}|\boldsymbol{y})}{Z_{\ell(1)}(\boldsymbol{y})}
\end{equation*}
\State Compute the expectation $ \boldsymbol{m}_{\ell(1)}$ of $\pi_{\ell(1)}^{\boldsymbol{y}}(\boldsymbol{\theta})$ w.r.t. $\pi_0(\boldsymbol{\theta})$
\State Compute the covariance $\boldsymbol{C}_{\ell(1)}$ of $\pi_{\ell(1)}^{\boldsymbol{y}}(\boldsymbol{\theta})$ w.r.t. $\pi_0(\boldsymbol{\theta})$

\State \Return $\pi_{\ell(1)}^{\boldsymbol{y}}(\boldsymbol{\theta}), \boldsymbol{m}_{\ell(1)}, \boldsymbol{C}_{\ell(1)}$
\EndProcedure
\end{algorithmic}
\end{algorithm}

\subsection{Level $\ell(j)$ with $j \geq 2$}
\subsubsection{Gaussian approximation} \label{subsubsec:gaussian_approx}
At levels $\ell(j)$ with j $\geq 2$, we sequentially update the prior density such that the previous level posterior, $\pi_{\ell(j - 1)}^{\boldsymbol{y}}(\boldsymbol{\theta})$, is used as the prior; we detail the sequential update in \cref{subsubsec:level_geq2_2}.
To be able to construct sparse grid approximations w.r.t. $\pi_{\ell(j - 1)}^{\boldsymbol{y}}(\boldsymbol{\theta})$, the underlying stochastic space needs to have a separable density (recall Assumption A1).
However, $\pi_{\ell(j - 1)}^{\boldsymbol{y}}(\boldsymbol{\theta})$ is usually not separable.
Therefore, we approximate $\pi_{\ell(j - 1)}^{\boldsymbol{y}}(\boldsymbol{\theta})$ with a density $\widehat{\pi}_{\ell(j - 1)}^{\boldsymbol{y}}(\boldsymbol{\theta})$ that allows us to obtain the required product structure.
In this paper, we approximate $\pi_{\ell(j - 1)}^{\boldsymbol{y}}(\boldsymbol{\theta})$ with the Gaussian density $\widehat{\pi}_{\ell(j - 1)}^{\boldsymbol{y}}(\boldsymbol{\theta})$ defined as
\begin{equation}\label{eq:gaussian_approx}
\widehat{\pi}_{\ell(j - 1)}^{\boldsymbol{y}}(\boldsymbol{\theta}) := \mathrm{N}(\boldsymbol{m}_{\ell(j - 1)}, \boldsymbol{C}_{\ell(j - 1)}),
\end{equation}
where $\boldsymbol{m}_{\ell(j - 1)}$ and $\boldsymbol{C}_{\ell(j - 1)}$ are the expectation and covariance matrix of $\pi_{\ell(j - 1)}^{\boldsymbol{y}}(\boldsymbol{\theta})$.

In most cases $\boldsymbol{C}_{\ell(j - 1)}$ is not diagonal, i.e., $\widehat{\pi}_{\ell(j - 1)}^{\boldsymbol{y}}(\boldsymbol{\theta})$ does not have a product structure.
Nevertheless, from the spectral decomposition $\boldsymbol{C}_{\ell(j - 1)} = VDV^{-1}$, we have $\boldsymbol{C}_{\ell(j - 1)}^{1/2} = VD^{1/2}V^{-1}$.
We arrive at
\begin{equation}\label{eq:gaussian_approx_mapping}
\boldsymbol{\theta} = T_{\ell(j - 1)}(\boldsymbol{\zeta}) := \boldsymbol{m}_{\ell(j - 1)} + \boldsymbol{C}_{\ell(j - 1)}^{1/2} \boldsymbol{\zeta} \Rightarrow \boldsymbol{\theta} \sim \mathrm{N}(\boldsymbol{m}_{\ell(j - 1)}, \boldsymbol{C}_{\ell(j - 1)}),
\end{equation}
where $\boldsymbol{\zeta}$ is a standard Gaussian random variable, i.e., $\boldsymbol{\zeta} \sim \mathrm{N}(\boldsymbol{0}, I)$.

Formula \cref{eq:gaussian_approx_mapping} allows to write a general multivariate Gaussian random variable with correlated components as a mapping of a standard multivariate Gaussian random variable, which has the desired product structure since the components of $\boldsymbol{\zeta}$ are uncorrelated and thus independent. 
In our context, we use \cref{eq:gaussian_approx_mapping} as follows.
We first generate $1$D (L)-Leja points weighted w.r.t. the standard normal density, that is, $w(\theta) :=  \exp{(-\theta^2/2)}/{\sqrt{2 \pi}}$ in \cref{eq:weighted_leja_points}.
Moreover, since the maximization defined in \cref{eq:weighted_leja_points} is typically performed over a compact domain, we consider $X_i := \mathbb{R} \approx [-4, 4]$. 
For quadrature, we compute $1$D quadrature weights w.r.t. normalized Hermite polynomials.
We extend these constructions to $N_\sto$ dimensions via tensorization and employ \cref{eq:gaussian_approx_mapping} to obtain the desired weighted (L)-Leja points.
Note that $T_{\ell(j - 1)}$ in \cref{eq:gaussian_approx_mapping} can be seen as an \emph{affine transport map} (see \cite{Ma16}).

\subsubsection{Level update on tensor domain} \label{subsubsec:level_geq2_2}
We assume that $\pi_{\ell(j - 1)}^{\boldsymbol{y}}(\boldsymbol{\theta})$ for $j \geq 2$ is not separable, hence we employ the Gaussian approximation \cref{eq:gaussian_approx_mapping} for all adaptive sparse grid operations; if $\pi_{\ell(j - 1)}^{\boldsymbol{y}}(\boldsymbol{\theta})$ is separable, everything that follows is computed directly using $\pi_{\ell(j - 1)}^{\boldsymbol{y}}(\boldsymbol{\theta})$ for all levels greater than two.
Thus, we sequentially update the prior in Bayes' formula \cref{eq:posterior} such that we reuse the Gaussian approximation of the posterior from the previous level, i.e.,
\begin{equation}\label{eq:seq_updated_bayes}
\begin{split}
\pi_{j}^{\boldsymbol{y}}(\boldsymbol{\theta}) 
&:= \frac{L_{j}(\boldsymbol{\theta}|\boldsymbol{y})\pi_0(\boldsymbol{\theta})}{Z_{j}(\boldsymbol{y})} = \frac{L_{j}(\boldsymbol{\theta}|\boldsymbol{y}) \pi_0(\boldsymbol{\theta})}{Z_{j}(\boldsymbol{y})} \frac{L_{j - 1}(\boldsymbol{\theta}|\boldsymbol{y})}{L_{j - 1}(\boldsymbol{\theta}|\boldsymbol{y})} \frac{Z_{j - 1}(\boldsymbol{y})}{Z_{j - 1}(\boldsymbol{y})}  \\
&= \frac{L_{j-1}(\boldsymbol{\theta}|\boldsymbol{y}) \pi_0(\boldsymbol{\theta})}{Z_{j-1}(\boldsymbol{y})} \frac{L_{j}(\boldsymbol{\theta}|\boldsymbol{y})}{L_{j - 1}(\boldsymbol{\theta}|\boldsymbol{y})} \frac{Z_{j - 1}(\boldsymbol{y})}{Z_{j}(\boldsymbol{y})} 
=  
\frac{\pi_{j - 1}^{\boldsymbol{y}}(\boldsymbol{\theta}) \frac{L_{j}(\boldsymbol{\theta}|\boldsymbol{y})}{L_{j - 1}(\boldsymbol{\theta}|\boldsymbol{y})}}{\frac{Z_{j}(\boldsymbol{y})}{Z_{j-1}(\boldsymbol{y})}} \\ 
& \approx 
\frac{\pi_{\ell(j - 1)}^{\boldsymbol{y}}(\boldsymbol{\theta}) L_{\delta j}(\boldsymbol{\theta}|\boldsymbol{y})}{Z_{\delta j}(\boldsymbol{y})}
=  
\frac{\widehat{\pi}_{\ell(j - 1)}^{\boldsymbol{y}}(\boldsymbol{\theta}) L_{\delta j}(\boldsymbol{\theta}|\boldsymbol{y}) \frac{\pi_{\ell(j - 1)}^{\boldsymbol{y}}(\boldsymbol{\theta})}{\widehat{\pi}_{\ell(j - 1)}^{\boldsymbol{y}}(\boldsymbol{\theta})} }{Z_{\delta j}(\boldsymbol{y})},
\end{split}
\end{equation}
where $L_{\delta j}(\boldsymbol{y}):= L_{j}(\boldsymbol{y})/L_{j-1}(\boldsymbol{y})$ and $Z_{\delta j}(\boldsymbol{y}):= Z_{j}(\boldsymbol{y})/Z_{j-1}(\boldsymbol{y})$.
Note that in \cref{eq:seq_updated_bayes} we correct the bias introduced by the Gaussian approximation of the posterior from the previous level, $\widehat{\pi}_{\ell(j - 1)}^{\boldsymbol{y}}$, with the ratio $\pi_{\ell(j - 1)}^{\boldsymbol{y}}/\widehat{\pi}_{\ell(j - 1)}^{\boldsymbol{y}}$.
 
First, we construct an adaptive sparse grid interpolation surrogate $\Phi_{\ell(\delta j)}(\boldsymbol{\theta};\boldsymbol{y})$ of $\Phi_{\delta j}(\boldsymbol{\theta};\boldsymbol{y})$ w.r.t. the density $\widehat{\pi}_{\ell(j - 1)}^{\boldsymbol{y}}(\boldsymbol{\theta})$ because it holds
\begin{equation*}
 L_{\delta j}(\boldsymbol{\theta}|\boldsymbol{y}) := \frac{\exp{(-\Phi_{{j}}(\boldsymbol{\theta};\boldsymbol{y}))}}{\exp{(-\Phi_{j - 1}(\boldsymbol{\theta};\boldsymbol{y}))}} = \exp{(-\Phi_{\delta j}(\boldsymbol{\theta};\boldsymbol{y}))}.
\end{equation*}
Recall that the mapping $T_{\ell(j - 1)}$ defined in \cref{eq:gaussian_approx_mapping} allows us to use adaptive sparse grid interpolation w.r.t. the Gaussian approximation $\widehat{\pi}_{\ell(j - 1)}^{\boldsymbol{y}}(\boldsymbol{\theta})$.
To construct the surrogate for the potential function $\Phi_{\delta j}(\boldsymbol{\theta};\boldsymbol{y})$ we employ $T_{\ell(j - 1)}$ and obtain
\begin{equation*}
\Phi_{\delta j}(\boldsymbol{\theta}; \boldsymbol{y}) \approx \Phi_{\ell({\delta j})}(T_{\ell(j - 1)}(\boldsymbol{\zeta}); \boldsymbol{y}).
\end{equation*}
$\Phi_{\ell({\delta j})}$ gives the approximation $L_{\ell({\delta j})}(T_{\ell(j - 1)}(\boldsymbol{\zeta})|\boldsymbol{y}) := \exp{(-\Phi_{\ell(\delta j)}(T_{\ell(j - 1)}(\boldsymbol{\zeta});\boldsymbol{y}))}$.

To evaluate the ratio of evidences $Z_{\delta j}(\boldsymbol{y})$ we make use of \cref{eq:seq_updated_bayes}, i.e.,
\begin{equation} \label{eq:evidence_ml}
\begin{split}
Z_{\delta j}(\boldsymbol{y}) 
&= \int_{\boldsymbol{X}} \widehat{\pi}_{j - 1}^{\boldsymbol{y}}(\boldsymbol{\theta}) L_{\delta j}(\boldsymbol{\theta}|\boldsymbol{y}) \frac{\pi_{j - 1}^{\boldsymbol{y}}(\boldsymbol{\theta})}{\widehat{\pi}_{j - 1}^{\boldsymbol{y}}(\boldsymbol{\theta})} \mathrm{d}\boldsymbol{\theta} \\
&\approx
\int_{\boldsymbol{X}} \widehat{\pi}_{\ell(j - 1)}^{\boldsymbol{y}}(T_{\ell(j - 1)}(\boldsymbol{\zeta})) \cdot L_{\ell(\delta j)}(T_{\ell(j - 1)}(\boldsymbol{\zeta})|\boldsymbol{y})) \cdot \frac{\pi_{\ell(j - 1)}^{\boldsymbol{y}}(T_{\ell(j - 1)}(\boldsymbol{\zeta}))}{\widehat{\pi}_{\ell(j - 1)}^{\boldsymbol{y}}(T_{\ell(j - 1)}(\boldsymbol{\zeta}))} \mathrm{d}\boldsymbol{\zeta} 
\end{split}
\end{equation}
and numerically integrate $\Big(L_{\ell({\delta j})} \pi_{\ell(j - 1)}^{\boldsymbol{y}}/\widehat{\pi}_{\ell(j - 1)}^{\boldsymbol{y}}\Big) \circ T_{\ell(j - 1)}$ w.r.t. the density $\widehat{\pi}_{\ell(j - 1)}^{\boldsymbol{y}}(\boldsymbol{\theta})$ via adaptive sparse grid quadrature to obtain the approximation $Z_{\ell(\delta j)}(\boldsymbol{y}) \approx Z_{\delta j}(\boldsymbol{y})$.
Thus, at each level $\ell(j)$ with $j \geq 2$ we obtain the posterior approximation
\begin{equation*}
\pi_{j}^{\boldsymbol{y}}(\boldsymbol{\theta}) \approx \pi_{\ell(j)}^{\boldsymbol{y}}(\boldsymbol{\theta}) := Z^{-1}_{\ell(\delta j)}(\boldsymbol{y}) \pi_{\ell(j - 1)}^{\boldsymbol{y}}(T_{\ell(j - 1)}(\boldsymbol{\zeta})) L_{\ell(\delta j)}(T_{\ell(j - 1)}(\boldsymbol{\zeta})|\boldsymbol{y})).
\end{equation*}

For all other quadrature computations, we proceed analogously to \cref{eq:evidence_ml}.
To simplify the notation, denote $R_{\ell(j - 1)} := \pi_{\ell(j - 1)}^{\boldsymbol{y}}/\widehat{\pi}_{\ell(j - 1)}^{\boldsymbol{y}}$.
Given an integrable function $g(\boldsymbol{\theta})$, we integrate $\int_{\boldsymbol{X}} g(\boldsymbol{\theta}) \pi_{j}^{\boldsymbol{y}}(\boldsymbol{\theta}) \mathrm{d}\boldsymbol{\theta}$, which reads as
\begin{equation*}
\begin{split}
\int_{\boldsymbol{X}} g(T_{\ell(j - 1)}(\boldsymbol{\zeta})) L_{\ell(\delta j)}(T_{\ell(j - 1)}(\boldsymbol{\zeta})|\boldsymbol{y})) R_{\ell(j - 1)}(T_{\ell(j - 1)}(\boldsymbol{\zeta})) \widehat{\pi}_{\ell(j - 1)}^{\boldsymbol{y}}(T_{\ell(j - 1)}(\boldsymbol{\zeta})) \mathrm{d}\boldsymbol{\zeta},
\end{split}
\end{equation*}
via adaptive sparse grid quadrature w.r.t. $\widehat{\pi}_{\ell(j - 1)}^{\boldsymbol{y}}$.
The above formula is used to assess the expectation and covariance matrix of $\pi_{j}^{\boldsymbol{y}}(\boldsymbol{\theta})$.
Moreover, at level $J$ of our approach, we integrate the QoI $g \approx g_{\ell(J)}$ w.r.t. $\widehat{\pi}_{\ell(J)}^{\boldsymbol{y}}(\boldsymbol{\theta})$.
In this way, we perform the multilevel decomposition implicitly, different from standard multilevel methods in which the QoI is assessed explicitly via telescoping sums.
Note that at the end of our multilevel algorithm we also obtain a surrogate for the posterior density which can be further used, for example, in an uncertainty propagation setting.

\begin{figure}
    \centering
  \tikzstyle{decision} = [diamond, draw, fill=black!8, 
    text width=5em, text badly centered, node distance=4cm, inner sep=0pt]
    \tikzstyle{blocklarge} = [rectangle, draw, fill=black!0, 
    text width=10em, text centered, rounded corners, minimum height=4em]
    \tikzstyle{blocklarger} = [rectangle, draw, fill=black!0, 
    text width=15em, text centered, rounded corners, minimum height=4em]
\tikzstyle{block} = [rectangle, draw, fill=black!0,
    text width=7.5em, text centered, rounded corners, minimum height=4em]
\tikzstyle{blocksmall} = [rectangle, draw, fill=black!0,
    text width=4em, text centered, rounded corners, minimum height=2.4em]
\tikzstyle{line} = [draw, -latex', line width = 0.8pt]
\tikzstyle{cloud} = [draw, rectangle, rounded corners, fill=black!15, node distance=4.6cm,
    minimum height=2em,text width=11em]
\tikzstyle{cloudsmall} = [draw, rectangle, rounded corners, fill=black!15, node distance=4.2cm,
    minimum height=2em,text width=4.5em]
    
\begin{tikzpicture}[node distance = 2cm, auto,scale=0.6, every node/.style={scale=0.6}]]
    \node [blocklarge] (init) {Use adaptive sparse grids to compute $\Phi_{\ell(1)}, \pi^{\boldsymbol{y}}_{\ell(1)}, \boldsymbol{m}_{\ell(1)}$,  $\boldsymbol{C}_{\ell(1)}$ w.r.t. the density $\pi_0$};
    \node [cloud, left of=init] (expert) {Input $=$ \\$J, h, tol, K_{\max}, \Phi, \pi_0, g$};
    %\node [cloud, right of=init] (system) {system};
    \node [blocksmall, below of=init, node distance=2cm] (identify) {$j \leftarrow 2$};
	%\node [blocklarge, below of=identify] (for_init) {Update density (4.1)};
    \node [decision, below of=identify, node distance=2.4cm] (evaluate) {$\pi^{\boldsymbol{y}}_{\ell(j-1)}$ has product structure?};
   
    \node [blocklarge, below of=evaluate, node distance=3cm] (decide) {$\widehat{\pi}^{\boldsymbol{y}}_{\ell(j-1)}\leftarrow{\pi}^{\boldsymbol{y}}_{\ell(j-1)}$};
    \node [blocklarge, right of=decide, node distance=4.4cm] (Gauss) {Gaussian \\ approximation to obtain $\widehat{\pi}^{\boldsymbol{y}}_{\ell(j-1)}$ and mapping $T_{\ell(j-1)}$};
    \node [blocklarge, below of=decide, node distance=2.4cm] (Compute_ratio_potential) {Use adaptive sparse grids to compute $\Phi_{\ell(\delta j)}, Z_{\ell(\delta j)}, \boldsymbol{m}_{\ell(j)}$, $\boldsymbol{C}_{\ell(j)}$ w.r.t. the density $\widehat{\pi}^{\boldsymbol{y}}_{\ell(j-1)}$};
    \node [blocklarger, below of=Compute_ratio_potential, node distance=2.4cm] (approx_post) { ${\pi}^{\boldsymbol{y}}_{\ell(j)} \leftarrow \pi^{\boldsymbol{y}}_{\ell(j-1)}L_{\ell(\delta j)}/Z_{\ell(\delta j)}$};
        \node [decision, below of=approx_post, node distance=2.4cm] (end_for) {Is $j < J$?};
        \node [cloudsmall, right of=end_for, node distance=4.5cm] (end) {Compute integral of QoI $g_{\ell{(J)}}$};
 \node [block, left of=Compute_ratio_potential, node distance=5.2cm] (update) {$j \leftarrow j+1$};

    % Draw edges
    \path [line] (init) -- (identify);
  % \path [line] (identify) -- (for_init);
  %  \path [line] (for_init) -- (evaluate);
  \path [line] (identify) -- (evaluate);
    \path [line] (evaluate) -- node {yes}(decide);
    \path [line] (end_for) -| node [near start] {yes} (update);
  
	\path [line] (update) |- (evaluate);
    \path [line] (evaluate) -| node [near start] {no}(Gauss);
    \path [line] (end_for) -- node {no}(end);

    \path [line] (Gauss) |- (Compute_ratio_potential);
    \path [line] (decide) -- (Compute_ratio_potential);
    \path [line] (Compute_ratio_potential) -- (approx_post);
    \path [line] (approx_post) -- (end_for);
    \path [line] (expert) -- (init);
    %\path [line,dashed] (system) -- (init);
    %\path [line,dashed] (system) |- (evaluate);
\end{tikzpicture}
    \caption{Flowchart of the multilevel adaptive sparse Leja algorithm. The goal is to approximate the posterior density $\pi^{\boldsymbol{y}}$ and $\mathbb{E}_{\pi^{\boldsymbol y}}[g]$ for an output QoI $g$ at level $J$ given the prior density $\pi_0$.}
    \label{fig:flowchart}
\end{figure}
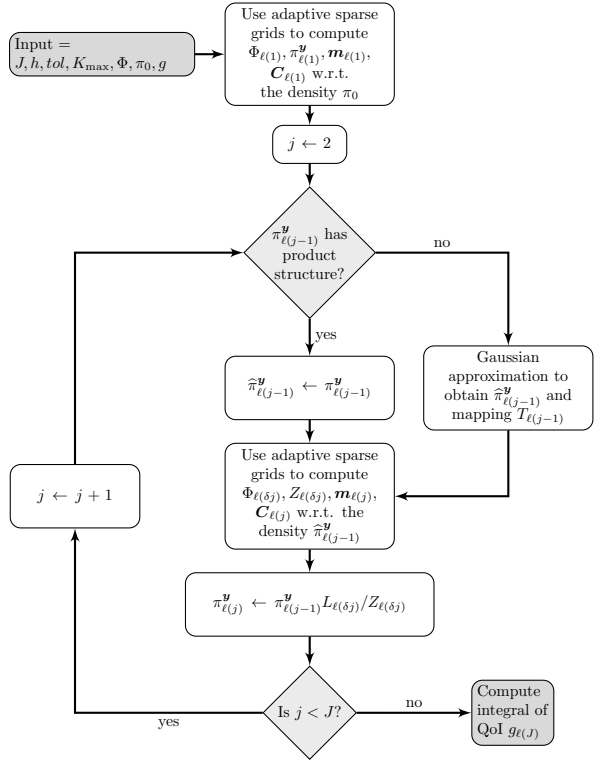

We summarize the steps for levels greater than two in  \cref{algo:our_approach_remainder}.
The first three inputs are the number of levels, $J$, the sequence of mesh sizes, $\boldsymbol{h}$, and $\boldsymbol{tol}$, which comprises the tolerances for adaptive sparse grid interpolation and quadrature at all levels.
The next input denotes the maximum reachable levels for the adaptive algorithms, $\boldsymbol{K}_{\mathrm{max}} := (K_{\mathrm{max}}^{\mathrm{qu}}, K_{\mathrm{max}}^{\mathrm{in}})$. 
Finally, $\pi_0$ is the prior density, $\Phi(\boldsymbol{\theta};\boldsymbol{y})$ is the potential function and $g$ is the QoI.
We combine \cref{algo:our_approach_step_1,algo:our_approach_remainder} and depict all steps in our proposed multilevel approach in \cref{fig:flowchart}.

\begin{algorithm}
\caption{Multilevel Adaptive Sparse Leja Algorithm}\label{algo:our_approach_remainder}
\begin{algorithmic}[1]
\Procedure{MultilevelAdaptSparseLeja}{$J, \boldsymbol{h}, \boldsymbol{tol},\boldsymbol{K}_{\mathrm{max}}, \Phi(\boldsymbol{\theta};\boldsymbol{y}), \pi_0(\boldsymbol{\theta}), g$}
\State Use \cref{algo:our_approach_step_1} to obtain
\begin{equation*}
\pi_{\ell(1)}^{\boldsymbol{y}}(\boldsymbol{\theta}), \boldsymbol{m}_{\ell(1)}, \boldsymbol{C}_{\ell(1)} = \mathrm{Level1SparseLeja}(h_1, tol^{\mathrm{in}}_1, tol^{\mathrm{qu}}_1, \boldsymbol{K}_{\mathrm{max}}, \Phi, \pi_0)
\end{equation*} 
\For{$j\gets 2, J$}
\State Construct the Gaussian approximation \cref{eq:gaussian_approx} and the mapping \cref{eq:gaussian_approx_mapping} \label{gaussian_approx}
\begin{equation*}
\widehat{\pi}_{\ell(j - 1)}^{\boldsymbol{y}}(\boldsymbol{\theta}) := \mathrm{N}(\boldsymbol{m}_{\ell(j - 1)}, \boldsymbol{C}_{\ell(j - 1)}), \quad T_{\ell(j - 1)}(\boldsymbol{\zeta}) := \boldsymbol{m}_{\ell(j - 1)} + \boldsymbol{C}_{\ell(j - 1)}^{1/2} \boldsymbol{\zeta}
\end{equation*}
\State Compute the potentials ratio surrogate via adaptive interpolation
\begin{equation*}
\Phi_{\ell(\delta j)}(\boldsymbol{\theta};\boldsymbol{y}) =
\mathrm{\texttt{AdaptSGInterp}}(tol_{j}^{\mathrm{in}}, K_{\mathrm{max}}^{\mathrm{in}}, \Phi_{\delta j} \circ T_{\ell(j - 1)}(\boldsymbol{\zeta}), \widehat{\pi}_{\ell(j - 1)}^{\boldsymbol{y}})
\end{equation*} 
\State Construct $L_{\ell({\delta j})}(T_{\ell(j - 1)}(\boldsymbol{\zeta})|\boldsymbol{y}) := \exp{(-\Phi_{\ell(\delta j)}(T_{\ell(j - 1)}(\boldsymbol{\zeta});\boldsymbol{y}))}$
\State Compute the evidence ratio via adaptive quadrature
\begin{equation*}
Z_{\ell(\delta j)}(\boldsymbol{y}) = \mathrm{\texttt{AdaptSGQuad}}(tol_{j}^{\mathrm{qu}}, K_{\mathrm{max}}^{\mathrm{qu}}, L_{\ell(\delta j)}, \widehat{\pi}_{\ell(j - 1)}^{\boldsymbol{y}})
\end{equation*}
\State Compute the updated posterior \label{ml_end}
\begin{equation*}
\pi_{\ell(j)}^{\boldsymbol{y}}(\boldsymbol{\theta}) := \frac{\pi_{\ell(j - 1)}^{\boldsymbol{y}}(\boldsymbol{\theta}) L_{\ell({\delta j})}(\boldsymbol{\theta}|\boldsymbol{y})}{Z_{\ell(\delta j)}(\boldsymbol{y})}
\end{equation*}
\State Compute the expectation $ \boldsymbol{m}_{\ell(j)}$ of $\pi_{\ell(j)}^{\boldsymbol{y}}(\boldsymbol{\theta})$ w.r.t. $\widehat{\pi}_{\ell(j - 1)}^{\boldsymbol{y}}(T_{\ell(j - 1)}(\boldsymbol{\zeta}))$
\State Compute the covariance $\boldsymbol{C}_{\ell(j)}$ of $\pi_{\ell(j)}^{\boldsymbol{y}}(\boldsymbol{\theta})$ w.r.t. $\widehat{\pi}_{\ell(j - 1)}^{\boldsymbol{y}}(T_{\ell(j - 1)}(\boldsymbol{\zeta}))$
\EndFor
\State Compute $g_{\ell(J)} \approx g$ and integrate w.r.t. $\pi_{\ell(J)}^{\boldsymbol y} \rightarrow \mathcal{I}_{\ell(J)}$ \label{qoi_computation}
\State \Return $\mathcal{I}_{\ell(J)}$
\EndProcedure
\end{algorithmic}
\end{algorithm}

\subsection{Computational cost} \label{subsubsec:cost}
The largest computational effort in the proposed multilevel methodology is spent in finding the interpolation surrogates since this involves evaluations of the forward operator.
Thus, for $j = 1, 2, \ldots J$, let $C_{h_j}$ denote the cost of the evaluating once the forward model discretized using a mesh depending on $h_j$.
Additionally, let $N_{tol_{J - j + 1}^{\mathrm{in}}}$ denote the number of forward operator solves to achieve tolerance $tol_{J - j + 1}^{\mathrm{in}}$ for adaptive sparse interpolation.
Then, the \emph{total interpolation cost} of our approach reads
\begin{equation*}
\mathrm{cost_{MLSL}^{\mathrm{in}}} = \sum_{j = 1}^{J - 1} (N_{tol_{J - j + 1}^{\mathrm{in}}} + N_{tol_{J - j}^{\mathrm{in}}})C_{h_j} + N_{tol_{1}^{\mathrm{in}}} C_{h_J}.
\end{equation*}
We obtain the above number because for each level except the last one, the same potential function and thus the same forward model enters two different likelihood ratios (see \cref{eq:seq_updated_bayes}). 
However, because we update the prior as the level increases, we expect the number of forward model solves to decrease significantly with the level.

All other costs are due to computations depending on the interpolation surrogates. 
These costs are however insignificant compared to the evaluation cost of a computationally expensive forward operator.

% adaptivity
\section{Dimension adaptivity with sparse grids} \label{sec:dim_adapt}
In this section we discuss the construction of the multiindex set $\mathcal{K}$ used in the sparse grid approximations defined in \cref{subsec:sparse_grids} via adaptive refinement.
For interpolation we consider a standard adaptive strategy as well as an enhanced approach that employs directional variance information.
For quadrature we employ a standard adaptive strategy.
In the following, our notation is similar to \cite{Fa18b}.
 
\subsection{Standard dimension-adaptive interpolation and quadrature} \label{subsubsec:std_adapt} 
Adaptive refinement is preferred especially when the underlying problem has a richer structure, such as anisotropic coupling of the input parameters or lower intrinsic dimensionality -- which is typically the case in most problems (see, e.g., \cite{CM13, Fa18b, Fa18, Wi16}).
The standard strategy is based on the dimension-adaptive algorithm of \cite{GG03, He03}.
The algorithm is described, e.g., in \cite{CM13, GG03, NJ14}.
We summarize only the basic idea below.
$\mathcal{K} = \{ \boldsymbol{1} \}$ initially.
Each refinement step is performed using the following principle: if a current multiindex contributes significantly to the approximation, its adjacent neighbours are likely to contribute as well.
Therefore, the forward neighbours of the multiindex with the largest contribution are added to $\mathcal{K}$ provided that $\mathcal{K}$ remains admissible.
The contribution of each $\boldsymbol{k}$ is assessed via a \emph{refinement indicator} $\epsilon(\boldsymbol{\boldsymbol{k}})$, whose choice
has a crucial impact on the performance of the adaptive algorithm.

We define 
\begin{equation}\label{eq:std_error_indicator}
\epsilon(\boldsymbol{k}) := s^{\mathrm{op}}(\boldsymbol{\Delta}^{\mathrm{op}}_{\boldsymbol{k}}[f^{\boldsymbol{N_\sto}}])/\delta N_{\boldsymbol{k}},
\end{equation}
where $s^{\mathrm{op}}$ is a function depending on the multivariate surplus $\boldsymbol{\Delta}_{\boldsymbol{k}}^{\mathrm{op}}[f^{\boldsymbol{N_\sto}}]$ and $\delta N_{\boldsymbol{k}}$ is the number model evaluations needed to assess $\boldsymbol{\Delta}_{\boldsymbol{k}}^{\mathrm{op}}[f^{\boldsymbol{N_\sto}}]$.
Note that $\delta N_{\boldsymbol{k}}$ penalizes subspaces with a large number of points. 

For sparse grid quadrature, we consider
\begin{equation*}
s^{\mathrm{qu}}(\boldsymbol{\Delta}^{\mathrm{qu}}_{\boldsymbol{k}}[f^{\boldsymbol{N_\sto}}]) := \| \boldsymbol{\Delta}^{\mathrm{qu}}_{\boldsymbol{k}}[f^{\boldsymbol{N_\sto}}] \|_{L^1} = |\boldsymbol{\Delta}^{\mathrm{qu}}_{\boldsymbol{k}}[f^{\boldsymbol{N_\sto}}]|,
\end{equation*}
which is a surrogate for the local quadrature error. 
For sparse interpolation, we use
\begin{equation} \label{eq:adapt_interp_std}
s^{\mathrm{in}}(\boldsymbol{\Delta}^{\mathrm{in}}_{\boldsymbol{k}}[f^{\boldsymbol{N_\sto}}]) := \| \boldsymbol{\Delta}^{\mathrm{in}}_{\boldsymbol{k}}[f^{\boldsymbol{N_\sto}}] \|_{L^2}.
\end{equation}
As in \cite{Fa18b}, we employ $\| \boldsymbol{\Delta}^{\mathrm{in}}_{\boldsymbol{k}}[f^{\boldsymbol{N_\sto}}] \|_{L^2}$ in the standard refinement indicator \cref{eq:adapt_interp_std} because it yields the local variance contribution of the surplus to the total variance.
\subsection{Directional variance dimension-adaptive sparse interpolation}\label{subsubsec:mod_adapt}
 
Dimension adaptivity based on error indicators such as \cref{eq:std_error_indicator} does not inherently distinguish between the individual input parameters.
Since in most problems the input parameters are anisotropically coupled, we wish to exploit this structure and tune the adaptive process such that it
stops refining directions that are rendered unimportant.
This is particularly important in our proposed approach since we update the prior starting with level $\ell(2)$ and thus we have updated information about the model's stochastic input parameters.
To this end, for sparse interpolation, we enhance the standard adaptive strategy such that we additionally compute a global measure of importance of each input parameter via \emph{total directional variances}, and we stop refining the directions having insignificant total directional variances. 

%In \cite{Su08} it was shown that the spectral coefficients can be used to analytically compute Sobol' indices for global sensitivity analysis (see \cite{So00}) as well.
%The Sobol' indices quantify the contribution of uncertain input parameters and interaction thereof to the total resulted variance, which serves as a measure of output uncertainty. 
%They are computed as ratios between directional variances, which quantify the contribution of each input or input interactions to the total variance, and the total resulted variance. 
%For further details on Sobol' indices, we refer to \cite{Fo13, So00, Su08}.

To enhance the standard adaptive approach for interpolation, we proceed analogously to \cite{Fa18b, Wi16} and perform a Sobol' decomposition \cite{So00} of the active set $\mathcal{A}$ to obtain directional variance surpluses, that is, 
\begin{equation*}
\Big \lVert \sum_{\boldsymbol{k} \in \mathcal{A}} \boldsymbol{\Delta}_{\boldsymbol{k}}^{\mathrm{in}}[f^{\boldsymbol{N_\sto}}] \Big \rVert_{L^2}^2 = \Delta V_{\mathcal{A}}^{0} + \sum_{i=1}^{N_\sto} \Delta V_{\mathcal{A}}^{i} + \Delta V_{\mathcal{A}}^{\mathrm{int}},
\end{equation*}
where $\Delta V_{\mathcal{A}}^{0} := \gamma_{\boldsymbol{0}}^2$ refers to the expectation contribution, 
\begin{equation*}
\Delta V_{\mathcal{A}}^{i} := \sum_{\boldsymbol{\ell} \in \mathcal{J}_{i}} \Delta \gamma_{\boldsymbol{\ell}}^2, \quad i=1, \ldots, N_\sto,
\end{equation*}
where $\mathcal{J}_{i} = \{ \boldsymbol{0} < \boldsymbol{\ell} \leq \boldsymbol{I}_{\boldsymbol{k}}: \boldsymbol{\ell}_i \neq 0 \land \boldsymbol{\ell}_j = 0,  \forall j \neq i \}$, are surplus contributions to all individual variances, and $\Delta V_{\mathcal{A}}^{\mathrm{int}}:= \sum_{\boldsymbol{m} \in \mathcal{J}_{\mathrm{int}}} \Delta \gamma_{\boldsymbol{m}}^2$ refers to the variance surplus due to all possible interactions, where 
$\mathcal{J}_{\mathrm{int}} = \bigcup_{i=1}^{N_\sto} \mathcal{J}_{i, \mathrm{int}}$, $\mathcal{J}_.{i, \mathrm{int}} = \{ \boldsymbol{0} < \boldsymbol{m} \leq \boldsymbol{I}_{\boldsymbol{k}}: \boldsymbol{m}_i \neq 0 \}$. 
Further, we compute \emph{total directional variance surpluses} as
\begin{equation*}
\Delta V_{\mathcal{A}}^{i, \mathrm{tot}} := \Delta V_{\mathcal{A}}^{i} + \Delta V_{\mathcal{A}}^{i, \mathrm{int}},
\end{equation*}
where $\Delta V_{\mathcal{A}}^{i, \mathrm{int}} := \sum_{\boldsymbol{m} \in \mathcal{J}_{i, \mathrm{int}}} \Delta \gamma_{\boldsymbol{m}}^2$ denotes the contribution due to all interactions involving direction $i$.
Note that $\Delta V_{\mathcal{A}}^{i, \mathrm{tot}}$ can be seen as a global measure of importance for each stochastic input: a large $\Delta V_{\mathcal{A}}^{i, \mathrm{tot}}$ implies that the $i$th parameter is significant from a stochastic perspective.
To this end, we prescribe $N_\sto$ user-defined \emph{directional tolerances} $\boldsymbol{\tau}^{\mathrm{in}} := (\tau_1^2, \tau_2^2, \ldots, \tau_{N_\sto}^2)$ and ascertain the importance of each input directions by comparing $\Delta V_{\mathcal{A}}^{i, \mathrm{tot}}$ with $\tau_i^2$ for $i = 1, 2, \ldots, N_\sto$. 
When the stochastic direction $i$ is rendered unimportant, we simply stop adding multiindices whose $i$th component exceeds the maximum $i$th index in the current multiindex set $\mathcal{K}$. 
In this way, the algorithm preferentially refines the most important directions, thus decreasing the overall computational cost.
When neither of the directional tolerances are met, the enhanced algorithm reduces to the standard approach in \cref{subsubsec:std_adapt}.

% results
\section{Numerical experiments} \label{sec:results}
In this section we present the numerical results obtained using our proposed multilevel approach for Bayesian inversion. 
%In the first test case, we focus on (L)-Leja quadrature in a simple inversion problem.
%In \cref{subsec:test_case_2}, we apply the proposed multilevel method and compare it with MCMC and a standard multilevel approach in a source inversion test case.
%To test the behaviour of our method in a problem with multimodal posterior, we investigate a source inversion problem with two sources in \cref{subsec:test_case_3}.
%Finally, we apply the proposed approach in a more complex and computationally more expensive numerical test in \cref{subsec:test_case_4}.

\subsection{Simple quadrature showcase} \label{subsec:test_case_1}
In this test case we investigate the behaviour of weighted (L)-Leja points in integration problems of the form
\begin{equation}\label{eq:int_post_quad}
\mathcal{I}(g) := \int g(\boldsymbol{\theta})\pi^{\boldsymbol{y}}(\boldsymbol{\theta})\mathrm{d}\boldsymbol{\theta},
\end{equation}
where $g(\boldsymbol{\theta})$ is an integrable function and $\pi^{\boldsymbol{y}}(\boldsymbol{\theta})$ is the posterior density; we outline the setup used to compute $\pi^{\boldsymbol{y}}(\boldsymbol{\theta})$ below. 
We assess \cref{eq:int_post_quad} via quadrature w.r.t. two different weight functions. 
In the first case, we employ a standard importance-sampling-based strategy (recall \cref{eq:ImportSampIdent}). 
Specifically, adaptive sparse grid quadrature w.r.t.the prior density, $\pi_{0}(\boldsymbol{\theta})$, with tolerance $tol^{\mathrm{qu}}_{\pi_{0}}$ is used:
\begin{equation}\label{eq:int_post_quad_prior}
\mathcal{I}(g) = 
\int g(\boldsymbol{\theta}) \frac{L(\boldsymbol{\theta}|\boldsymbol{y})\pi_0(\boldsymbol{\theta})}{Z(\boldsymbol{y})} \mathrm{d}\boldsymbol{\theta}
\approx Z_{N_{\mathrm{pr}}}^{-1}(\boldsymbol{y}) \big(\sum_{n=1}^{N_{\mathrm{pr}}} g(\boldsymbol{\theta}_{n, \mathrm{pr}}) L(\boldsymbol{\boldsymbol{\theta}}_{n, \mathrm{pr}}|\boldsymbol{y}) w_{n, \mathrm{pr}}\big),
\end{equation}
where $\{\boldsymbol{\theta}_{n, \mathrm{pr}}\}_{n=1}^{N_{\mathrm{pr}}}$ are (L)-Leja nodes computed w.r.t. $\pi_{0}(\boldsymbol{\theta})$ and
\begin{equation*}
Z_{N_{\mathrm{pr}}}(\boldsymbol{y}) = \sum_{n=1}^{N_{\mathrm{pr}}} \boldsymbol{\theta}_{n, \mathrm{pr}} L(\boldsymbol{\boldsymbol{\theta}}_{n, \mathrm{pr}}|\boldsymbol{y}) w_{n, \mathrm{pr}}.
\end{equation*}

In the second strategy, we compute \cref{eq:int_post_quad} using our proposed approach. 
We integrate \cref{eq:int_post_quad} numerically via adaptive sparse grid quadrature w.r.t.the Gaussian approximation $\widehat{\pi}^{\boldsymbol{y}}(\boldsymbol{\theta})$ of the posterior density (recall \cref{eq:gaussian_approx}), using a tolerance $tol^{\mathrm{qu}}_{\widehat{\pi}^{\boldsymbol{y}}}$: 
\begin{equation}\label{eq:int_post_quad_post}
\mathcal{I}(g) 
= \int g(\boldsymbol{\theta}) \frac{\pi^{\boldsymbol{y}}(\boldsymbol{\theta})}{\widehat{\pi}^{\boldsymbol{y}}(\boldsymbol{\theta})} \widehat{\pi}^{\boldsymbol{y}}(\boldsymbol{\theta}) \mathrm{d} \boldsymbol{\theta} \approx
\sum_{n=1}^{N_{\mathrm{post}}} g(T(\boldsymbol{\zeta}_{n, \mathrm{post}})) \frac{\pi^{\boldsymbol{y}}(T(\boldsymbol{\zeta}_{n, \mathrm{post}}))}{\widehat{\pi}^{\boldsymbol{y}}(T(\boldsymbol{\zeta}_{n, \mathrm{post}}))}w_{n, \mathrm{post}},
\end{equation}
where $\{\boldsymbol{\zeta}_{n, \mathrm{post}}\}_{n=1}^{N_{\mathrm{post}}}$ are (L)-Leja nodes computed w.r.t.the standard multivariate normal density, $N(\boldsymbol{0}, I)$, and $T(\boldsymbol{\zeta}) := \boldsymbol{m} + \boldsymbol{C}^{1/2} \boldsymbol{\zeta}$, where $\boldsymbol{m}$ and $\boldsymbol{C}$ are the expectation and covariance matrix associated with the density $\widehat{\pi}^{\boldsymbol{y}}(\boldsymbol{\theta})$.

\begin{figure}[htbp]
  \centering
  \includegraphics[width=0.75\textwidth]{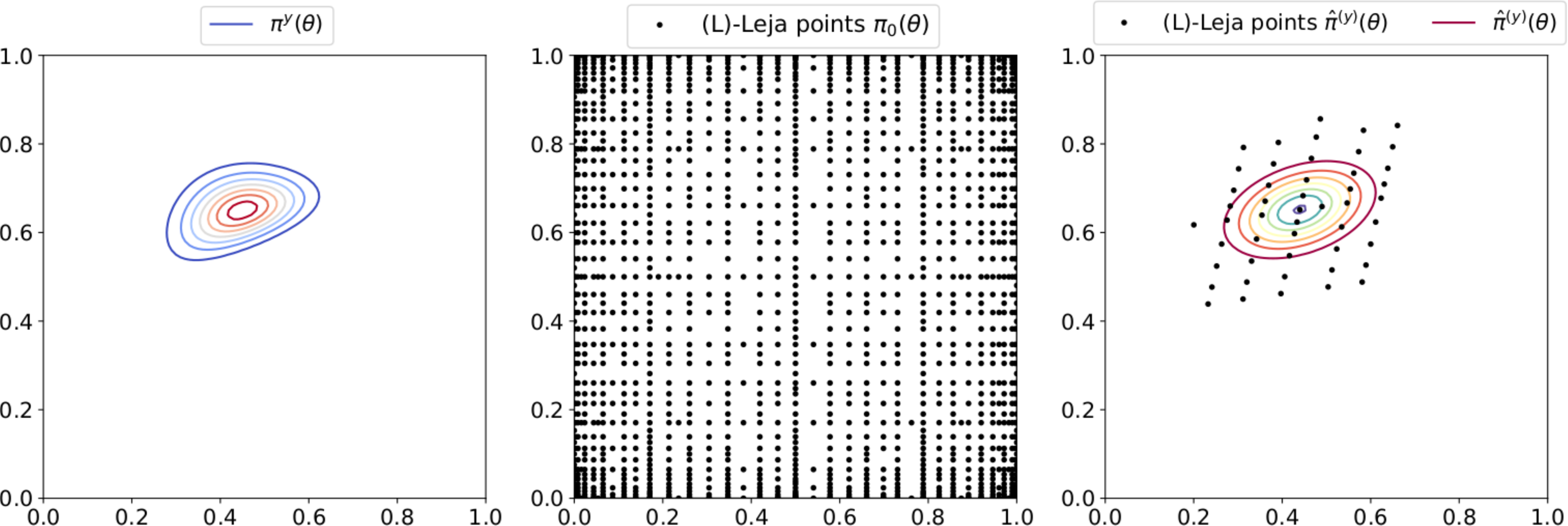}
  \caption{
	Left: Posterior density $\pi^{\boldsymbol{y}}(\boldsymbol{\theta})$.
   Center: (L)-Leja points computed w.r.t. the uniform prior density used in the integration problem \cref{eq:int_post_quad_prior}.
  Right: Gaussian approximation $\widehat{\pi}^{\boldsymbol{y}}(\boldsymbol{\theta})$ of the posterior and the associated weighted (L)-Leja points used in \cref{eq:int_post_quad_post}.}
  \label{fig:prob1D2D_post_comp}
\end{figure}

We consider the following forward model
\begin{equation*}
\mathcal{G}(\boldsymbol{\theta}) := \frac{A(\theta_1)}{(w(\theta_2) \pi)^2} \big( \sin{(w(\theta_2)\pi x)} - \sin{(w(\theta_2)\pi)}  x  \big),
\end{equation*} 
where $x \in [0, 1], A(\theta_1) = 20\theta_1 + 1$ and $w(\theta_2) = \theta_2 + 1.2$. 
We employ Bayesian inversion to infer $(\theta_1, \theta_2)$.
The prior is the uniform density in $[0, 1]^2$, i.e., $\pi_0 = U(0, 1)^2$.
The observation data $\boldsymbol{y}$ are generated synthetically using $(\theta_1, \theta_2)_{\mathrm{true}} = (0.45, 0.65)$.
We take $N_{obs} = 9$ measurements at locations $o_j = 0.1j, \quad j = 1, \ldots, 9$, assumed to be corrupted by additive Gaussian noise $\eta \sim \mathrm{N}(\boldsymbol{0}, 0.1^2I)$.
We depict the prior and posterior densities in the left figure in \cref{fig:prob1D2D_post_comp}.
Observe that the posterior is unimodal and non-symmetric, but it can be well approximated with a Gaussian density.

In the numerical experiments, we let $g(\boldsymbol{\theta}) := \exp{(-\theta_1 - \theta_2)}$ in \cref{eq:int_post_quad}.
We compute a reference solution using $3 \cdot 10^5$ Metropolis-Hastings MCMC (MH) samples obtained from a random walk Gaussian proposal with initial sample $\boldsymbol{\theta}_0 = (1, 1)$ and covariance matrix $C_{\mathrm{MH}} = 7 \cdot 10^{-3} I$. 
The acceptance rate is $44 \%$.
Additionally, we employ a tolerance $tol^{\mathrm{qu}}_{\pi_{0}} = 10^{-11}$ in \cref{eq:int_post_quad_prior} and a tolerance $tol^{\mathrm{qu}}_{\widehat{\pi}^{\boldsymbol{y}}} = 10^{-5}$ in \cref{eq:int_post_quad_post}.

The results are summarized in \cref{tab:res_test_case_1}.
The employed tolerances in the two (L)-Leja sparse grid quadrature approaches are sufficient to match four digits of the reference results.
However, integrating w.r.t. the prior requires $1603$ nodes, whereas our approach which uses the Gaussian approximation of the posterior as weight function requires only $49$ quadrature nodes, i.e., almost $33$ times fewer points.
This is because the support of the prior density, $\pi_0(\boldsymbol{\theta})$, is significantly larger than the support of the posterior: when integrating w.r.t. the prior density, the adaptive algorithm places a large number of quadrature points outside of the support of the posterior.
\begin{table}[tbhp]
{\footnotesize
  \caption{Results for the quadrature problem \cref{eq:int_post_quad} using
a reference MH solution with $3\cdot10^5$ samples, integration w.r.t. the prior density as in \cref{eq:int_post_quad_prior} and our proposed approach in which we integrate w.r.t. the Gaussian approximation of the posterior, as showed in \cref{eq:int_post_quad_post}.}\label{tab:res_test_case_1}
\begin{center}
  \begin{tabular}{|c|c|c|} \hline
   Method & No. quadrature points & Result \\ \hline
    MH & $3 \cdot 10^5$ &  0.33813 \\
    Integration w.r.t. $\pi_0$ & 1603 & 0.33813 \\
     Integration w.r.t. $\widehat{\pi}^{\boldsymbol{y}}(\boldsymbol{\theta})$ &  49 &  0.33811 \\ \hline
  \end{tabular}
\end{center}
}
\end{table}
We visualize the quadrature nodes corresponding to \cref{eq:int_post_quad_prior,eq:int_post_quad_post} in the center and right figures in \cref{fig:prob1D2D_post_comp}, respectively.
\begin{remark} \label{re:remark_quad}
Adaptive sparse grid quadrature w.r.t. an uninformative prior can sometimes stop early since the quadrature points will fall outside of the support of the integrated function, yielding null evaluations and thus null error indicators.
To overcome this, one could employ non-adaptive quadrature with a sufficiently large, a priori chosen number of nodes to cover the support of the integrated function.
\end{remark}

\subsection{Source inversion with one source in a 2D spatial domain} \label{subsec:test_case_2}
Consider a two dimensional Bayesian inverse problem in which the forward model $\mathcal{G}(\boldsymbol{\theta})$ is an elliptic PDE defined on $\Omega := [0, 1]^2$,
\begin{align} \label{eq:test_case_2}
-(u_{xx} + u_{yy})(x,y) &= A(\alpha) \exp{(-[(x - \theta_1)^2 + (y - \theta_2)^2)]/2\alpha^2)},  &(x, y) \in \Omega\\
u(x, y) &= 0, &(x, y) \in \partial \Omega, \nonumber
\end{align}
with $A(\alpha) = 5/(2\pi\alpha^2)$ and $\alpha = 0.2$.

The goal is to infer the coordinates $(\theta_1, \theta_2)$ of the source term in the right-hand side, i.e., we seek the solution to a source inversion problem.
We perform the multilevel Bayesian inversion as described in \cref{algo:our_approach_step_1,algo:our_approach_remainder} with three levels, i.e., $J = 3$.
Thus $j = 1, 2, 3$.
The employed multilevel setup is summarized in \cref{tab:ml_setup_test_case_2}.
Standard triangular finite elements (FEs) with mesh widths $h_j$ are used for spatial discretization.
To find surrogates for the potential function at each level in our proposed multilevel approach we employ both adaptive sparse interpolation variants summarized in \cref{sec:dim_adapt}.
Recall that for standard adaptive interpolation we have tolerances $tol^{\mathrm{in}}_1, tol^{\mathrm{in}}_2, tol^{\mathrm{in}}_3$ (see \cref{subsubsec:std_adapt}) whereas for directional variances-based adaptivity from \cref{subsubsec:mod_adapt} we additionally have the directional tolerances $\{\boldsymbol{\tau}^{\mathrm{in}}_j\}_{j=1}^3$.
We choose the FE mesh widths and adaptive interpolation tolerances such that the approximation errors are quantitatively similar.
At level $\ell(1)$ we combine $h_1$ with $tol^{\mathrm{in}}_3$ and $\boldsymbol{\tau}^{\mathrm{in}}_3$, at level $\ell(2)$, $h_2$ is combined with $tol^{\mathrm{in}}_2$ and $\boldsymbol{\tau}^{\mathrm{in}}_2$, and at level $\ell(3)$ we employ $h_3$ together with $tol^{\mathrm{in}}_1$ and $\boldsymbol{\tau}^{\mathrm{in}}_1$.
For adaptive sparse grid quadrature we employ small tolerances $tol^{\mathrm{qu}}_1, tol^{\mathrm{qu}}_2, tol^{\mathrm{qu}}_3$ to prevent the adaptive algorithm to stop too early especially when integrating w.r.t. the prior density (recall \cref{re:remark_quad}). 
\begin{table}[tbhp]
{\footnotesize
  \caption{Multilevel setup for the 2D inversion problem with forward model \cref{eq:test_case_2}.}\label{tab:ml_setup_test_case_2}
\begin{center}
  \begin{tabular}{|c|c|c|c|c|} \hline
   Level & $h$ & $tol^{\mathrm{in}}$ & $\boldsymbol{\tau}^{\mathrm{in}}$ & $tol^{\mathrm{qu}}$ \\ \hline
    $\ell(1)$ & $h_1 = \sqrt{2}/2^{4}$ & $tol^{\mathrm{in}}_3 = 10^{-5}$ & $\boldsymbol{\tau}^{\mathrm{in}}_3 = (10^{-7}, 10^{-7})$ & $tol^{\mathrm{qu}}_3 = 10^{-12}$ \\
    $\ell(2)$ & $h_2 = \sqrt{2}/2^{5}$ & $tol^{\mathrm{in}}_2 = 10^{-4}$ & $\boldsymbol{\tau}^{\mathrm{in}}_2 = (10^{-6}, 10^{-6})$ & $tol^{\mathrm{qu}}_2 = 10^{-11}$ \\
    $\ell(3)$ & $h_3 = \sqrt{2}/2^{6}$ & $tol^{\mathrm{in}}_1 = 10^{-3}$ & $\boldsymbol{\tau}^{\mathrm{in}}_1 = (10^{-5}, 10^{-5})$ & $tol^{\mathrm{qu}}_1 = 10^{-10}$ \\ \hline
  \end{tabular}
\end{center}
}
\end{table}

Since the source locations need to reside inside $\Omega$, the prior is the uniform density in $[0, 1]^2$, i.e., $\pi_0 = U(0, 1)^2$.
We consider 16 sensor locations at $(0.2i, 0.2j)$ for $i, j = 1, 2, 3, 4$. 
The measurements are obtained synthetically by discretizing the forward model on a finer mesh, i.e., $h = \sqrt{2}/2^7$ to avoid committing an ``inverse crime''.
Moreover, $\boldsymbol{\theta}_{\mathrm{true}} = (0.35, 0.65)$ and the additive Gaussian noise $\boldsymbol{\eta}\sim \text{N}(\mathbf{0}, 0.2^2 I)$.

The QoI is the  posterior mean $\mathbb{E}_{\pi^{\boldsymbol{y}}}[\boldsymbol{\theta}]$.
We compute a reference solution using $2 \cdot 10^5$ samples obtained from a random walk Metropolis-Hastings algorithm with Gaussian proposal having
covariance matrix $C_{\mathrm{MH}} = 4 \cdot 10^{-3} I$, started from $\boldsymbol{\theta}_0 = (1, 1)$.
The acceptance rate of the chain is $64 \%$.
To obtain a comprehensive overview of the accuracy and cost of our approach, we compare it with the standard three-level approach in which all adaptive sparse grid operations are performed w.r.t. the prior density.
Moreover, the QoI is assessed using the classical telescoping sum.
To simplify the notation, in the following we use the abbreviation $\mathrm{StdML}$ to refer to the standard multilevel approach.
$\mathrm{MLLejaStd}$ refers to our approach in which standard dimension-adaptive interpolation is used at each level and $\mathrm{MLLejaDV}$ refers to our approach combined with directional variance-based adaptive interpolation summarized in \cref{subsubsec:mod_adapt}.
The results are presented in \cref{tab:ml_res_test_case_2}.
Observe that all multilevel methods yield results very close to the reference estimate.
Thus, the two variants of our proposed approach, $\mathrm{MLLejaStd}$ and $\mathrm{MLLejaDV}$, are comparably accurate as the sampling-based and the standard multilevel solutions.
\begin{table}[tbhp]
{\footnotesize
  \caption{Comparison of estimates of $\mathbb{E}[\pi^{\boldsymbol{y}}(\boldsymbol{\theta})]$ for the source inversion problem with forward model \cref{eq:test_case_2}.
  We first compute a reference solution using $2 \cdot 10^5$ MH samples.
  Afterwards, we employ StdML and the two variants of our proposed multilevel approach, MLLejaStd and MLLejaDV.}\label{tab:ml_res_test_case_2}
\begin{center}
  \begin{tabular}{|c|c|} \hline
   Method & $\mathbb{E}_{\pi^{\boldsymbol{y}}}[\boldsymbol{\theta}]$  \\ \hline
    MH & $(0.3628, 0.6370)$ \\
    StdML & $(0.3631, 0.6368)$  \\
    MLLejaStd & $(0.3630, 0.6369)$  \\
    MLLejaDV & $(0.3630, 0.6369)$  \\ \hline
  \end{tabular}
\end{center}
}
\end{table}

\begin{figure}[htbp]
  \centering
  \includegraphics[width=0.75\textwidth]{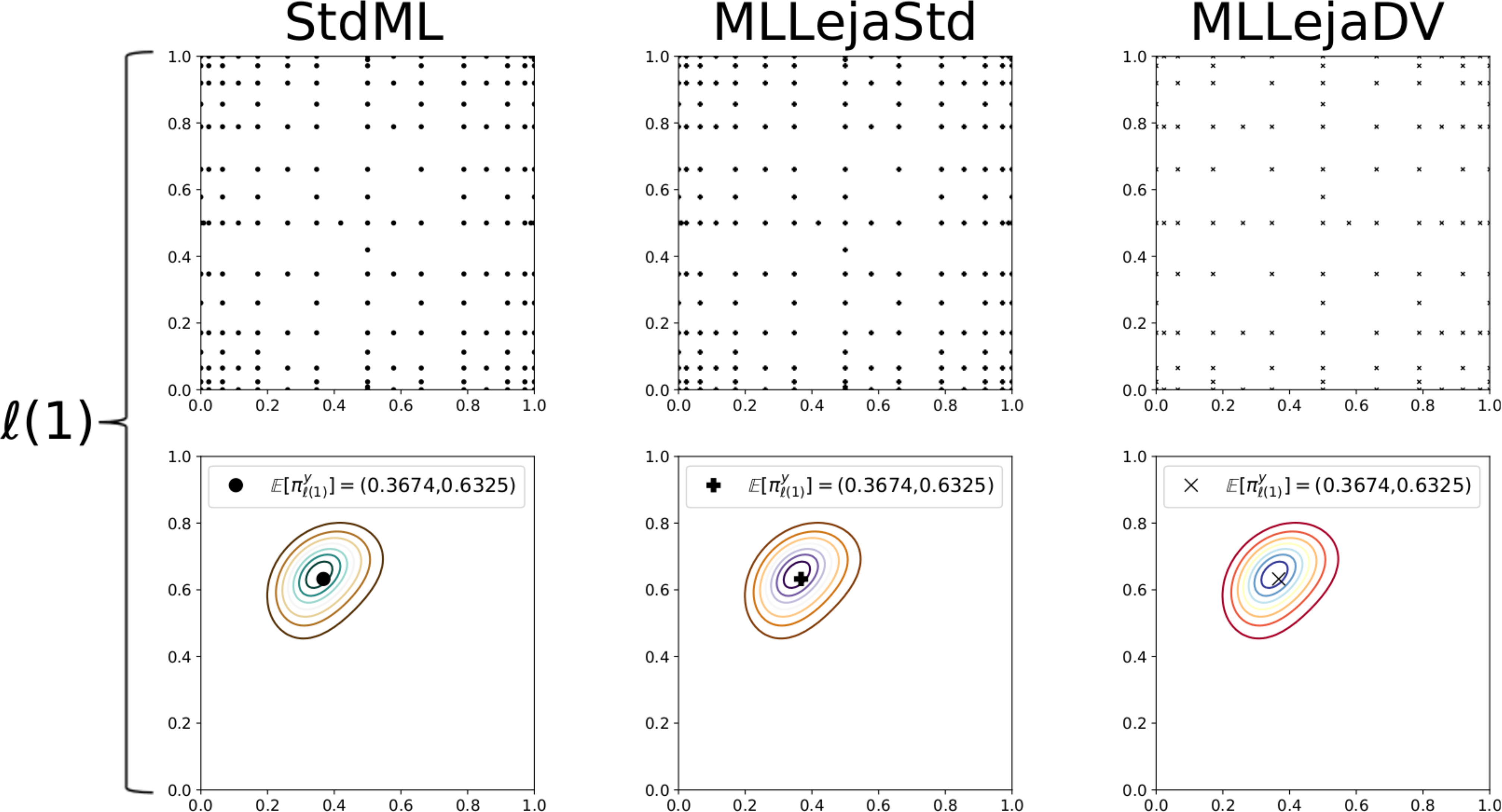}
  \includegraphics[width=0.75\textwidth]{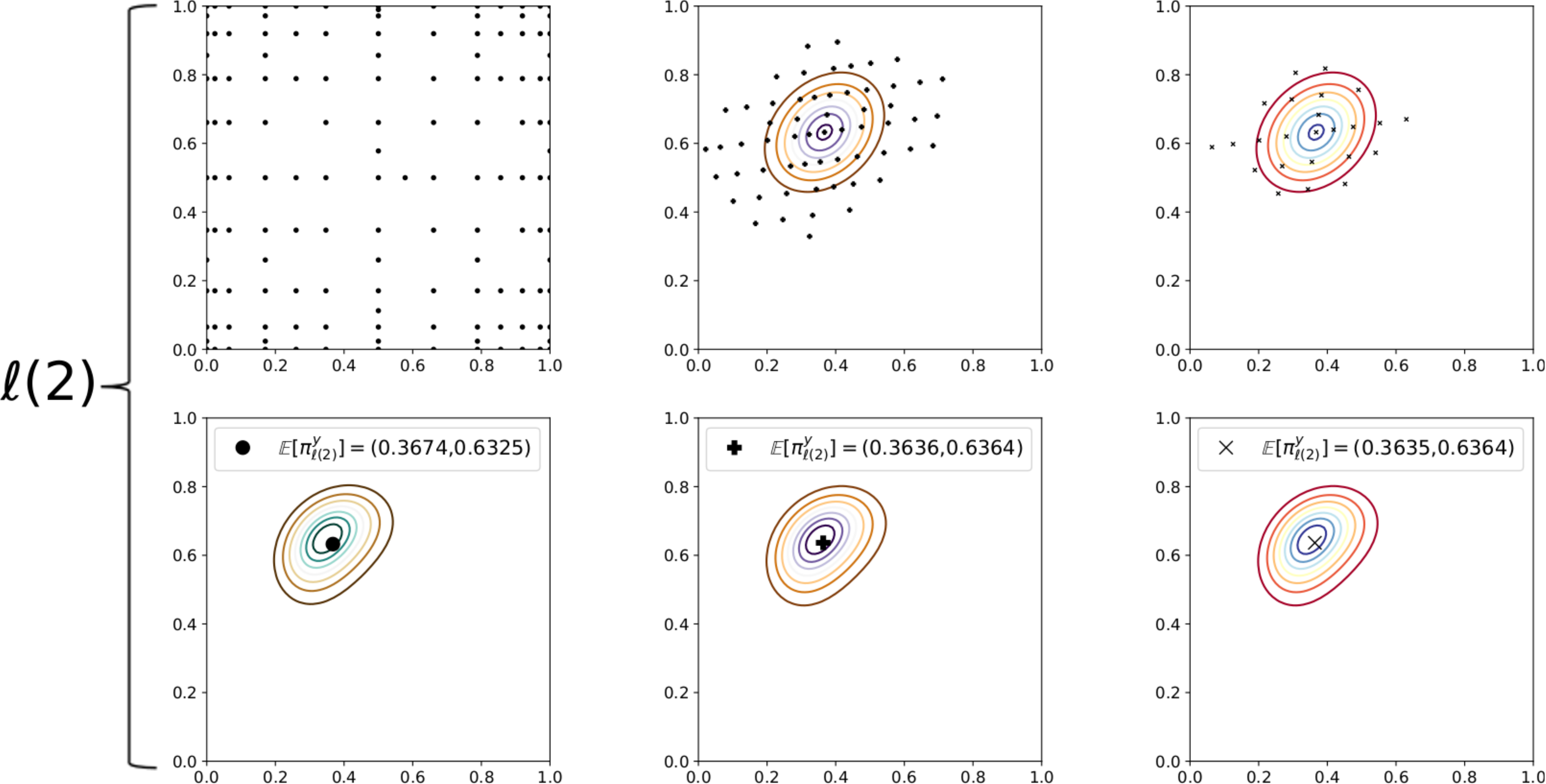}
  \includegraphics[width=0.75\textwidth]{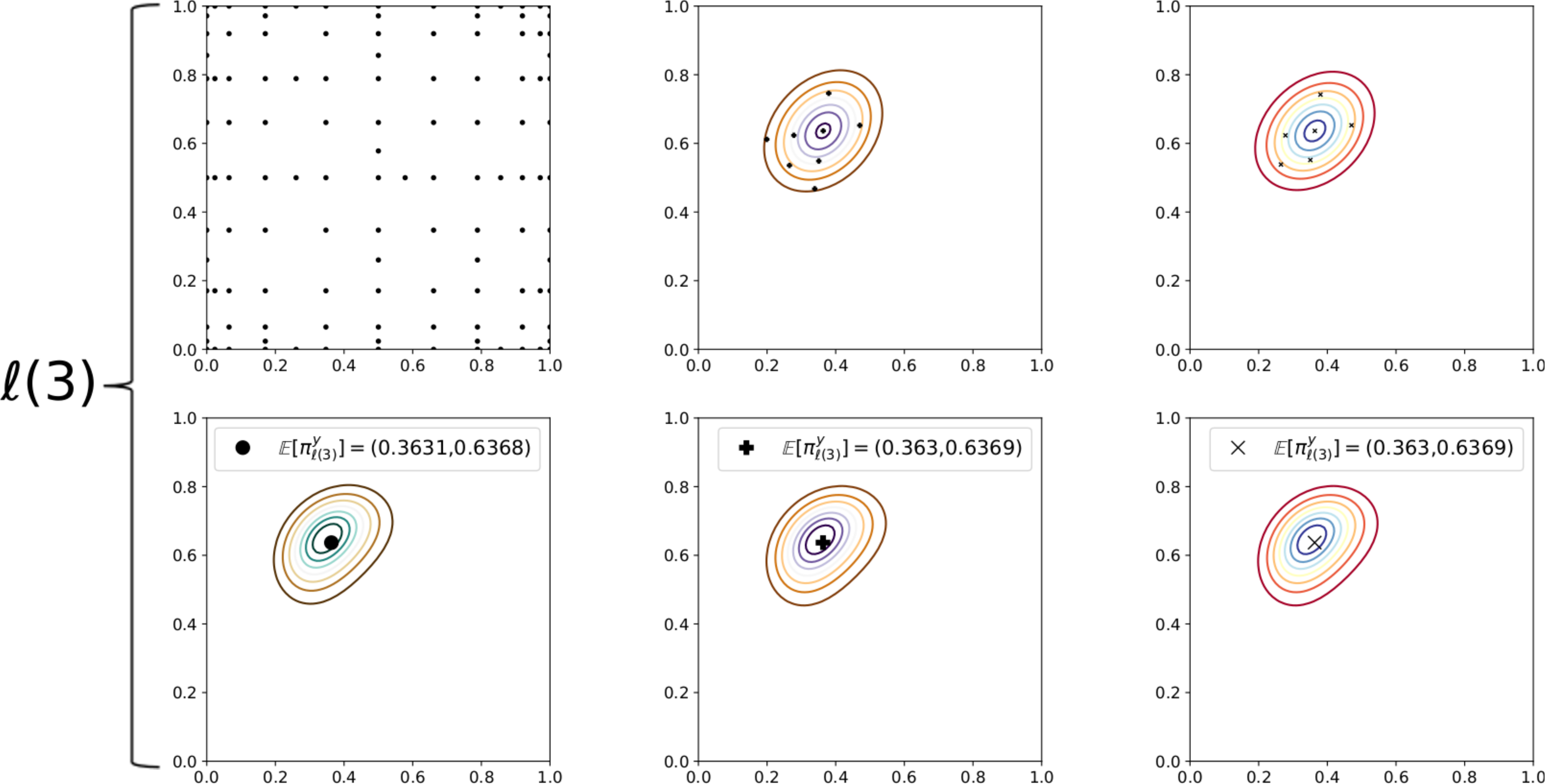}
  \caption{Results obtained using StdML (left), MLLejaStd (center) and MLLejaDV (right) to find the adaptive sparse grid interpolation surrogate for the potential function in the source inversion problem with forward model \cref{eq:test_case_2}.
  			At each of the three levels, in the top plots we depict the prior density and the corresponding weighted (L)-Leja points used to find the surrogate. 
  			At levels $\ell(2)$ and $\ell(3)$, the prior is the Gaussian approximation of the posterior from the posterior level.
  			In the bottom plots we depict the corresponding posterior density solution.}
  \label{fig:prob2D2D_1source_ml_densities}
\end{figure}

In \cref{fig:prob2D2D_1source_ml_densities}, we visualize the results for all employed multilevel methods as follows.
The left subplots show the results for $\mathrm{StdML}$, whereas the center and right subplots depict the results for $\mathrm{MLLejaStd}$ and $\mathrm{MLLejaDV}$ respectively.
Furthermore, at each level, the prior as well as the corresponding (L)-Leja points used to find the interpolation surrogate are visualized in the top part.
In the bottom plots, we depict the resulting posterior densities.
At level $\ell(1)$ we obtain the same three posteriors since the prior is the same in all cases.
However, starting with level $\ell(2)$ the sequential update of the prior in the proposed approach leads to significantly fewer interpolation points compared to $\mathrm{StdML}$, which places a large number weighted (L)-Leja points outside of the support of the corresponding posterior.
Moreover, comparing the two variants of the proposed approach, $\mathrm{MLLejaDV}$ requires fewer (L)-Leja points than $\mathrm{MLLejaStd}$.
This is because at both levels $\ell(2)$ and $\ell(3)$ in $\mathrm{MLLejaDV}$, the two total directional variances fall below the imposed tolerances.  
Thus $\mathrm{MLLejaDV}$ discovers and exploits more structure in the underlying approximation problem.

\begin{figure}[htbp]
  \centering
  \includegraphics[width=0.75\textwidth]{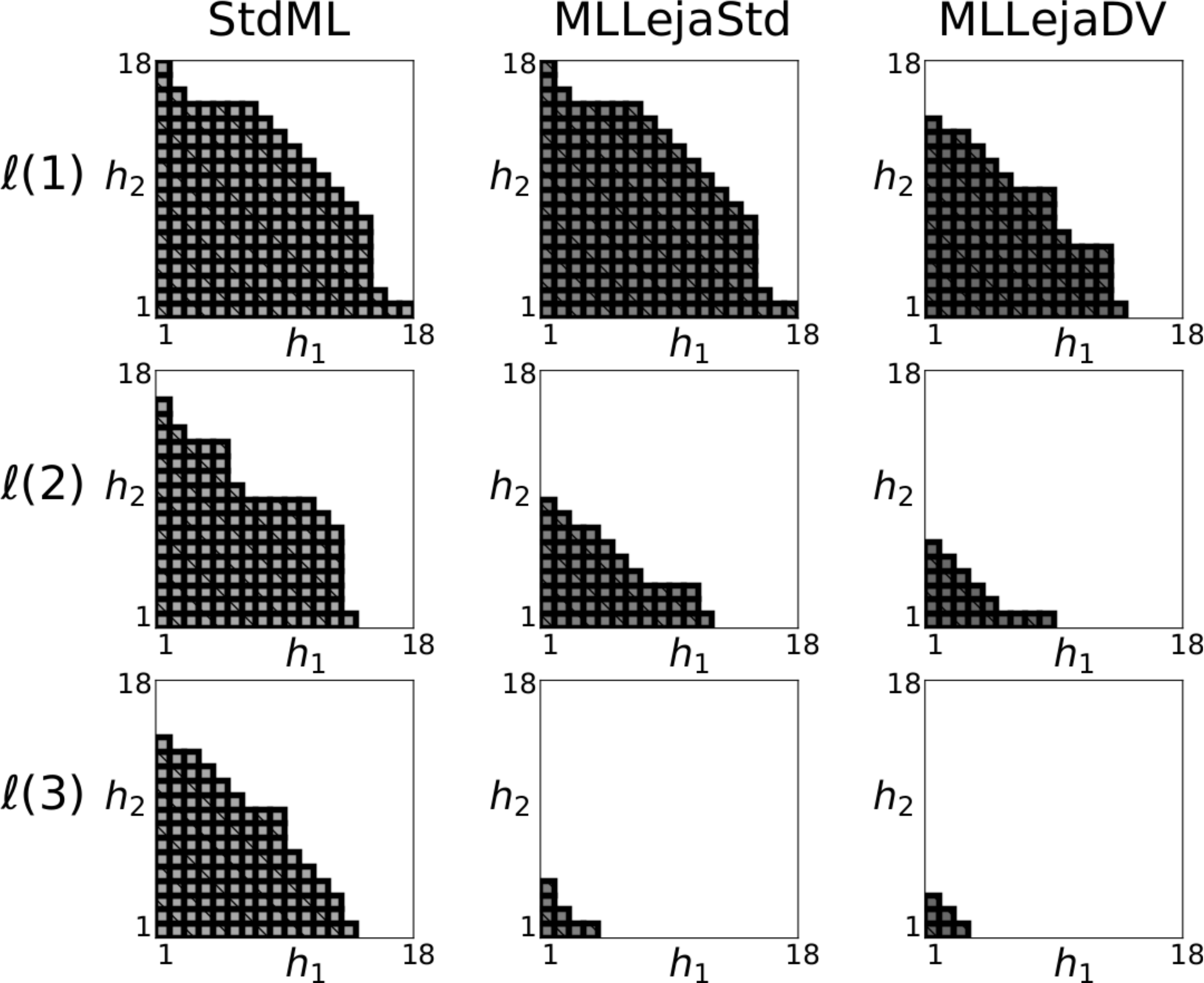}
  \caption{Multiindex sets corresponding to adaptive sparse grid interpolation in the source inversion problem \cref{eq:test_case_2} for StdML (left), MLLejaStd (center) and MLLejaDV (right).}
  \label{fig:prob2D2D_1source_ml_index_sets}
\end{figure}

We visualize the multiindex sets for the three multilevel variants in \cref{fig:prob2D2D_1source_ml_index_sets}.
At level $\ell(1)$ the multiindex sets corresponding to $\mathrm{StdML}$ and $\mathrm{MLLejaStd}$ are symmetric; this is because the two stochastic parameters have equal importance w.r.t. the prior density, which is the $2$D (symmetric) uniform density.
$\mathrm{MLLejaDV}$ leads to a smaller multiindex set since the directional variances $\boldsymbol{\tau}^{\mathrm{in}}_3$ fall below $(10^{-7}, 10^{-7})$, but there is no clear distinction between the two input directions as well. 
However, at level $\ell(2)$ we observe a different behaviour in the two variants of our proposed approach, $\mathrm{MLLejaStd}$ and $\mathrm{MLLejaDV}$.
Recall that in these two variants the prior density is the Gaussian approximation of the posterior density from level $\ell(1)$.
Computing the eigenvalues $(\lambda_1, \lambda_2)$ of its covariance matrix, we obtain $\lambda_1 = 0.0097$ and $\lambda_2 = 0.0055$. 
Therefore, the first direction is more important that the second, which is reflected in the two multiindex sets. 
On the other hand, the multiindex set for the standard approach remains symmetric because the prior density is unchanged.
Finally, at level $\ell(3)$ we see low-cardinality multiindex sets for both $\mathrm{MLLejaStd}$ and $\mathrm{MLLejaDV}$.
This is because at this stage we have an informative prior, thus the likelihood ratio is close $1$, which requires little approximation effort.
Hence going beyond level $\ell(3)$ is not necessary for our approach.

\begin{figure}[htbp]
  \centering
  \includegraphics[width=0.6\textwidth]{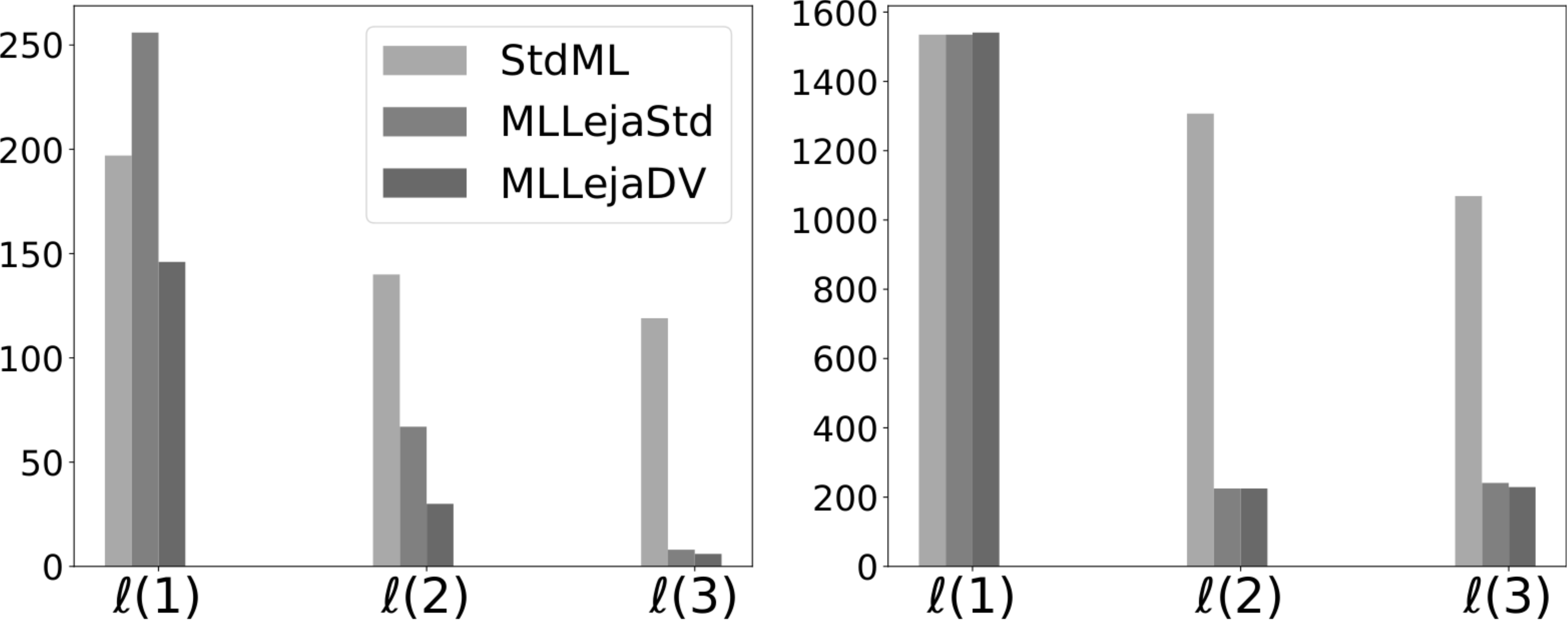}
  \caption{Left: total number of forward model evaluations needed in the adaptive sparse interpolation of the potential using the three multilevel variants in the source inversion problem \cref{eq:test_case_2}. 
  			Right: total number of quadrature nodes.}
  \label{fig:prob2D2D_1source_cost}
\end{figure}

The costs of all multilevel methods are visualized in \cref{fig:prob2D2D_1source_cost}.
The number of forward model evaluations needed to find the adaptive sparse grid surrogate of the potential function are shown on the left side.
In the right plot we depict the number of evaluations of the surrogate in all quadrature computations.
Note in all multilevel variants we need quadrature to assess the evidences and expectations at all three levels.
Additionally, in $\mathrm{MLLejaStd}$ and $\mathrm{MLLejaDV}$ we need to compute the covariance matrices at levels $\ell(1)$ and $\ell(2)$ as well, which are needed in the Gaussian approximation of the associated posteriors and affine mapping (recall \cref{eq:gaussian_approx,eq:gaussian_approx_mapping}).
However, since we integrate w.r.t. the same weight function, we keep all surrogate evaluations in a look-up table and reuse them whenever the same grid points are used for different evaluations.
We observe that at level $\ell(1)$ our proposed approach is slightly more expensive for interpolation which is due to the need to evaluate the FE solver on level $\ell(1)$ at level $\ell(2)$ as well: recall that at level $\ell(2)$ we construct a sparse grid surrogate for the ratio of potential functions. 
However, the increased cost is not significant since it involves evaluations of the coarsest FE solver, which are very fast.
Starting with level $\ell(2)$ we see significant cost savings for both interpolation and quadrature.
On the one hand, at level $\ell(2)$, $\mathrm{MLLejaStd}$ leads to about $2$ times fewer forward model evaluations for sparse grid interpolation and around $12.5$ fewer sparse grid quadrature evaluations.
Moreover, at level $\ell(3)$ we obtain about $15$ times fewer interpolation points and $7.5$ times fewer quadrature nodes.
On the other hand, $\mathrm{MLLejaDV}$ leads to about $5$ times fewer interpolation nodes and about $12.5$ times fewer quadrature evaluations.
Furthermore, we obtain $20$ times fewer interpolation nodes and $9.5$ times fewer quadrature nodes at level $\ell(3)$.
These results clearly show that updating the prior information in our multilevel approach for Bayesian Inversion leads to significant cost reduction in finding and evaluating sparse grid surrogates.

\subsection{Source inversion with two sources in a 2D spatial domain} \label{subsec:test_case_3}
For a more comprehensive overview of the proposed approach, we consider now a test case with multimodal observation data.
In particular, we consider another source inversion test case in which we use two sources to generate the data -- to have bimodal observation data -- and only one source to perform the Bayesian inference. 

The elliptic forward operator defined on $\Omega := [0, 1]^2$ reads:
\begin{align} \label{eq:test_case_3}
-(u_{xx} + u_{yy})(x,y) &= A(\alpha) \big(\exp{(-[(x - \theta_1)^2 + (y - \theta_2)^2)]/2\alpha^2)}  \nonumber & \\  
&\qquad + b \exp{(-[(x - \theta_3)^2 + (y - \theta_4)^2)]/2\alpha^2)} \big),  &(x, y) \in \Omega \\
u(x, y) &= 0, &(x, y) \in \partial \Omega, \nonumber
\end{align}
where $A(\alpha) = 5/(2\pi\alpha^2), \alpha = 0.15$ and the binary parameter $b = 1$ when generating the data and $b = 0$ when performing the inference.
Therefore, we are solving a source inversion similar to the one in \cref{subsec:test_case_2} but starting from bimodal data.

To generate the data we choose the locations of the two sources far apart, i.e., $(\theta_1, \theta_2)_{\mathrm{true}} = (0.15, 0.15)$ and $(\theta_3, \theta_4)_{\mathrm{true}} = (0.85, 0.85)$.
For Bayesian inference we employ StdML, MLLejaStd and MLLejaDV using three levels.
The multilevel setup is outlined in \cref{tab:ml_setup_test_case_3}.
\begin{table}[tbhp]
{\footnotesize
  \caption{Multilevel setup for the 2D inversion problem with forward model \cref{eq:test_case_3}.}\label{tab:ml_setup_test_case_3}
\begin{center}
  \begin{tabular}{|c|c|c|c|c|} \hline
   Level & $h$ & $tol^{\mathrm{in}}$ & $\boldsymbol{\tau}^{\mathrm{in}}$ & $tol^{\mathrm{qu}}$ \\ \hline
    $\ell(1)$ & $h_1 = \sqrt{2}/2^{5}$ & $tol^{\mathrm{in}}_3 = 10^{-6}$ & $\boldsymbol{\tau}^{\mathrm{in}}_3 = (10^{-8}, 10^{-8})$ & $tol^{\mathrm{qu}}_3 = 10^{-13}$ \\
    $\ell(2)$ & $h_2 = \sqrt{2}/2^{6}$ & $tol^{\mathrm{in}}_2 = 10^{-5}$ & $\boldsymbol{\tau}^{\mathrm{in}}_2 = (10^{-7}, 10^{-7})$ & $tol^{\mathrm{qu}}_2 = 10^{-12}$ \\
    $\ell(3)$ & $h_3 = \sqrt{2}/2^{7}$ & $tol^{\mathrm{in}}_1 = 10^{-4}$ & $\boldsymbol{\tau}^{\mathrm{in}}_1 = (10^{-6}, 10^{-6})$ & $tol^{\mathrm{qu}}_1 = 10^{-11}$ \\ \hline
  \end{tabular}
\end{center}
}
\end{table}

\begin{figure}[htbp]
  \centering
  \includegraphics[width=0.6\textwidth]{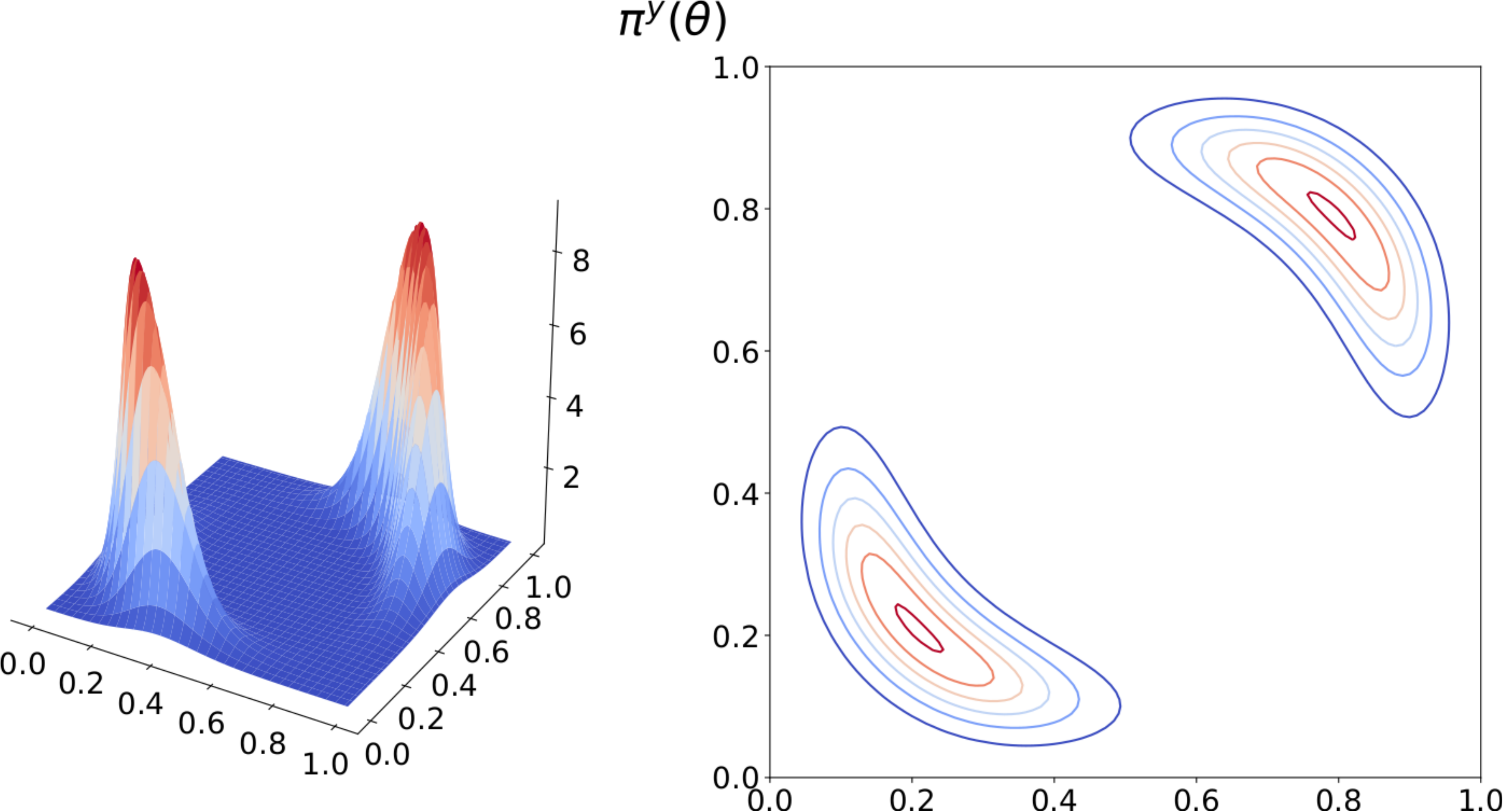}
  \caption{Posterior density for the source inversion problem with forward model \cref{eq:test_case_3}. 
  We used $50^2 = 2500$ Gauss-Legendre points to assess the evidence.}
  \label{fig:prob2D2D_2sources_ref_post}
\end{figure}

We visualize in \cref{fig:prob2D2D_2sources_ref_post} the bimodal posterior density obtained via standard Bayes' formula \cref{eq:posterior} for which we used $50^2 = 2500$ Gauss-Legendre points to assess the evidence.
Note that the two peaks are symmetric around $(0.5, 0.5)$.
Therefore, Bayesian inference using only once source in the forward model can yield, in the best case, a posterior density centered at $(0.5, 0.5)$.

The QoI is again $\mathbb{E}[\pi^{\boldsymbol{y}}(\boldsymbol{\theta})]$. 
In \cref{tab:ml_res_test_case_3} we show the obtained estimates.
First, a reference Metropolis-Hastings estimate with $2 \cdot 10^5$ samples  is computed using a random walk Gaussian proposal with initial sample $\boldsymbol{\theta}_0 = (1.0, 1.0)$ and covariance matrix $C_{\mathrm{MH}} = 10^{-1} I$.
The acceptance rate is $45 \%$.
We observe that the MH  and StdML solutions yield estimates close to the center, $(0.5, 0.5)$.
However, the estimates given by MLLejaStd and MLLejaDV are far away from this value.
\begin{table}[tbhp]
{\footnotesize
  \caption{Comparison of estimates of $\mathbb{E}[\pi^{\boldsymbol{y}}(\boldsymbol{\theta})]$ for the source inversion problem with forward model \cref{eq:test_case_3}.
  We first compute a reference solution using $2 \cdot 10^5$ MH samples.
  Afterwards, we employ StdML and the two variants of our proposed multilevel approach, MLLejaStd and MLLejaDV.}\label{tab:ml_res_test_case_3}
\begin{center}
  \begin{tabular}{|c|c|} \hline
   Method & $\mathbb{E}_{\pi^{\boldsymbol{y}}}[\boldsymbol{\theta}]$  \\ \hline
   MH & $(0.5032, 0.5068)$ \\
    StdML & $(0.5002, 0.5002)$  \\
    MLLejaStd & $(0.6688, 0.6548)$  \\
    MLLejaDV & $(0.6648, 0.6548)$  \\ \hline
  \end{tabular}
\end{center}
}
\end{table}

\begin{figure}[htbp]
  \centering
  \includegraphics[width=0.75\textwidth]{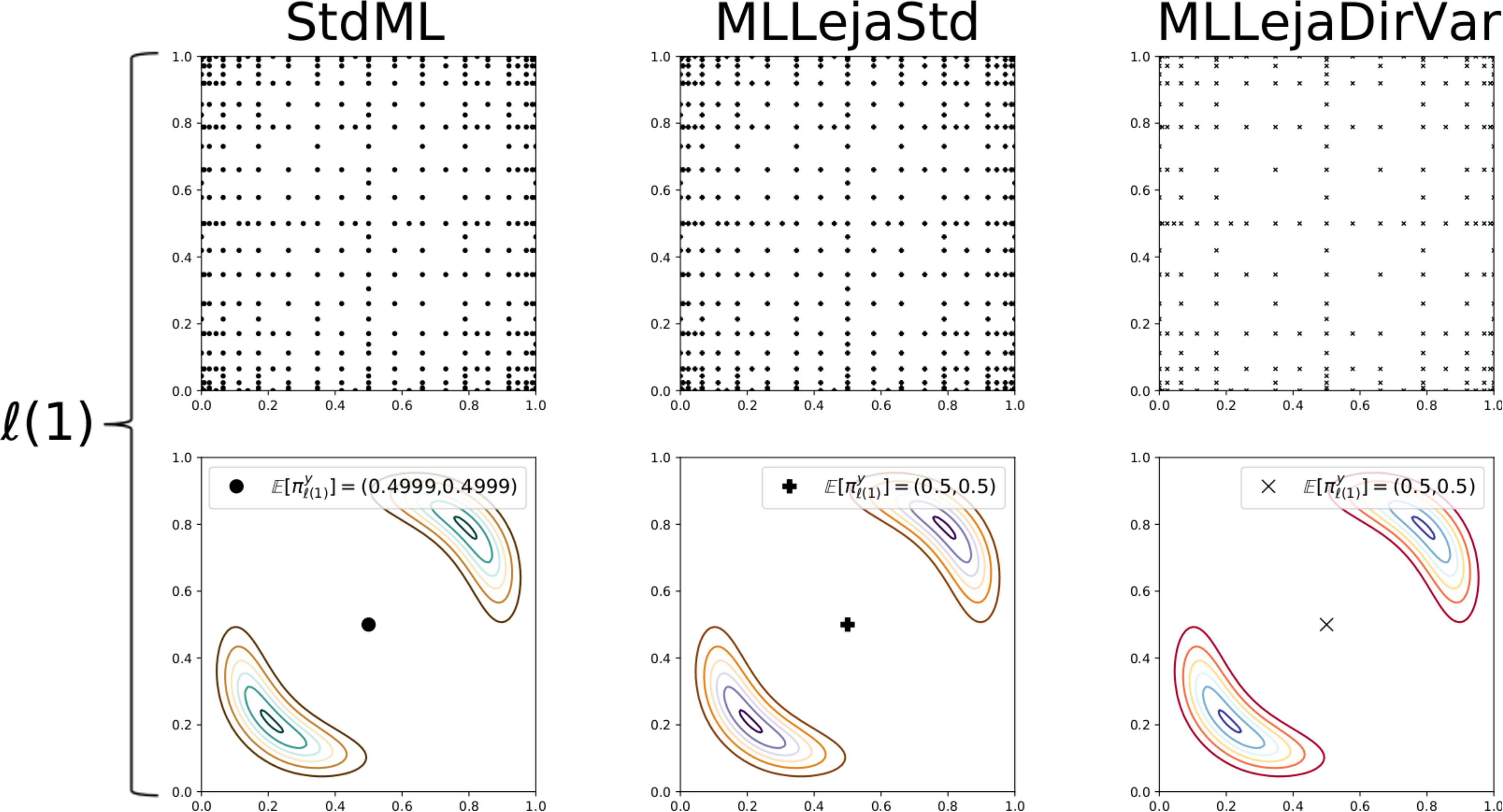}
  \includegraphics[width=0.75\textwidth]{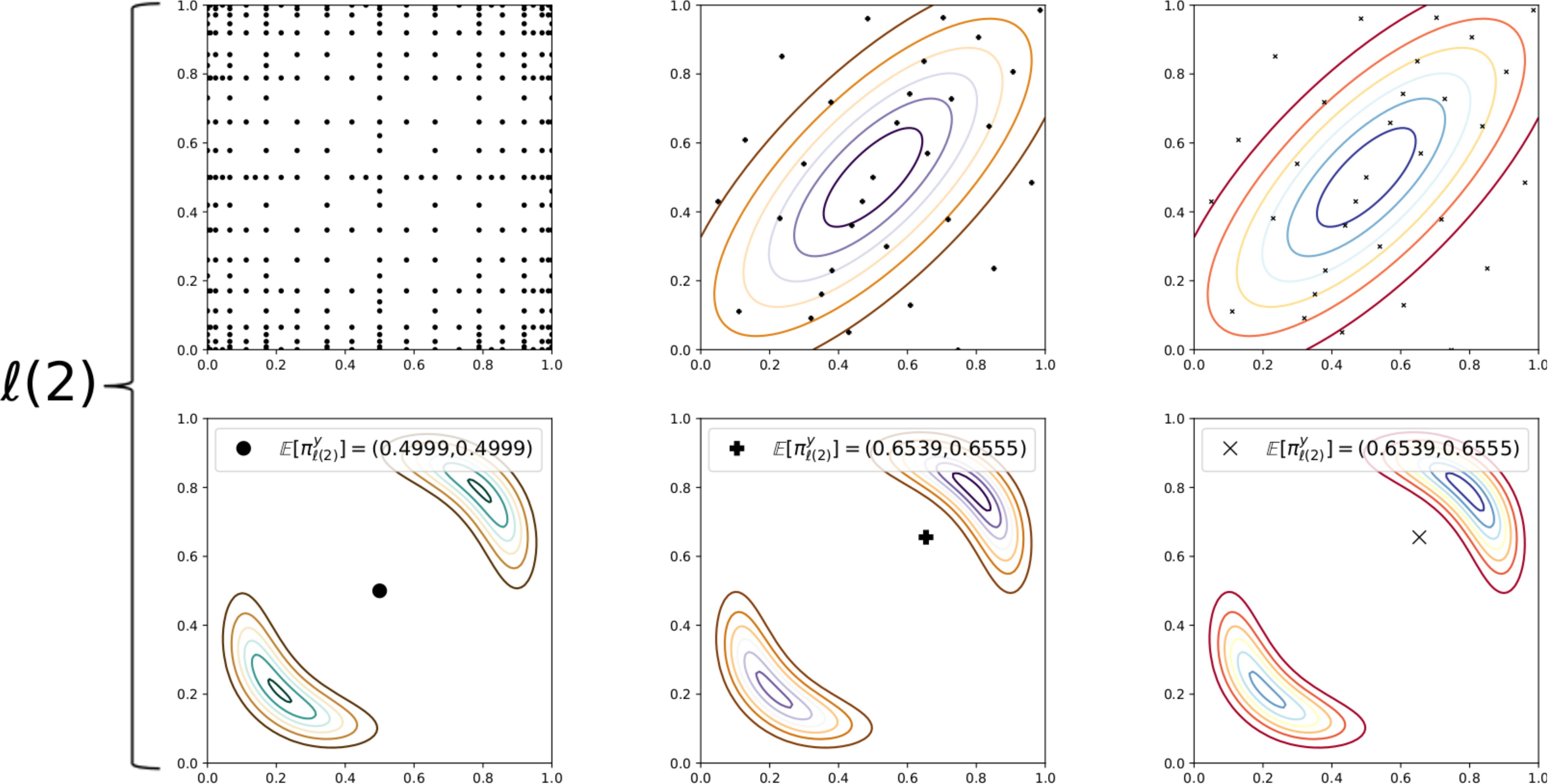}
  \includegraphics[width=0.75\textwidth]{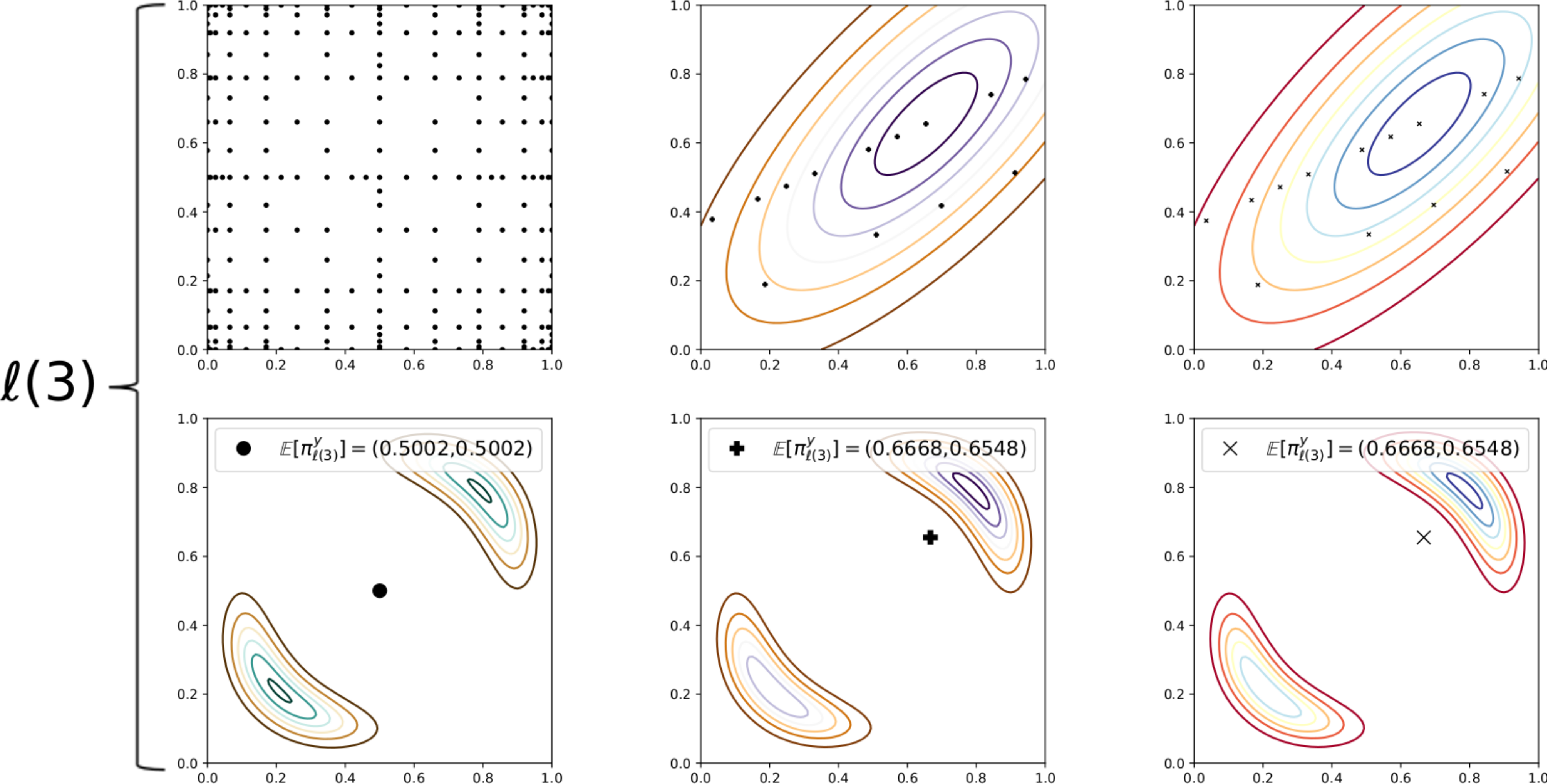}
  \caption{ Results obtained using StdML (left), MLLejaStd (center) and MLLejaDV (right) to find the adaptive sparse grid interpolation surrogate for the potential function in the source inversion problem with forward model \cref{eq:test_case_3}.
  			At each of the three levels, in the top plots we depict the prior density and the corresponding weighted (L)-Leja points used to find the surrogate. 
  			At levels $\ell(2)$ and $\ell(3)$, the prior is the Gaussian approximation of the posterior from the posterior level.
  			In the bottom plots we depict the corresponding posterior density solution.}
  \label{fig:prob2D2D_2sources_ml_densities}
\end{figure}
We depict in \cref{fig:prob2D2D_2sources_ml_densities} the prior and posterior density as well as the weighted (L)-Leja points used to construct the adaptive sparse grid interpolation surrogate of the potential function for all employed multilevel methods.
We observe that the Gaussian approximation used in our proposed approach at levels $\ell(2)$ and $\ell(3)$ is very spread and quite different from the approximated posterior.
Hence the bias-correcting ratio in quadrature operations, $\pi_{\ell(j)}^{\boldsymbol{y}}/\widehat{\pi}_{\ell(j)}^{\boldsymbol{y}}$ for $j = 2, 3$ (recall \cref{eq:seq_updated_bayes}) is different from $1$ in most regions of the domain.
The large variations of this ratio lead to large variations of the error indicators in adaptive sparse grid quadrature which prevent the algorithm to converge and hence to yield accurate estimates.
Therefore, the estimates of the expectation and covariance matrix of the posteriors from levels $\ell(1)$ and $\ell(2)$, and with that, the Gaussian approximations employed at levels $\ell(2)$ and $\ell(3)$, are inaccurate.
Note that the large spread of the Gaussian approximations leads to weighted (L)-Leja points outside of the domain of the uniform prior, which coincides with the domain of the forward operator, $\Omega$ (see \cref{eq:test_case_3}).
Whenever this happens, we impose the corresponding likelihood evaluation to be zero.
\subsection{Higher-dimensional problem in a 3D spatial domain} \label{subsec:test_case_4}
In the final test case, we apply the proposed approach in a more challenging and computationally more expensive problem.
We consider an elliptic forward model defined on $\Omega := [0, 1]^3$ with a permeability field projected onto a Fourier basis:
\begin{align} \label{eq:test_case_4}
-\nabla \cdot (k(x, y, z, \boldsymbol{\theta}) \nabla u(x,y,z)) &= f,   &(x, y, z) \in \Omega \\
u(x, y, z) &= 0,  &(x, y, z) \in \partial \Omega, \nonumber
\end{align}
where $f \equiv 5$ and  
\begin{equation} \label{eq:diff_test_case_4}
k(x, y, z, \boldsymbol{\theta}) :=  \exp{\Big(\sum_{n = 1}^8 s_n \theta_n \sin{(p_{n, 1} \pi x)} \sin{(p_{n, 2} \pi y)} \sin{(p_{n, 3} \pi z)} \Big) },
\end{equation}
where $s_1 = 0.1785$, $s_{2} = s_{3} = s_{4} = 0.1428$, $s_{5} = s_{6} = s_{7} = 0.1071, s_8 = 0.0714$ are normalized scaling factors, i.e., $\sum_{n=1}^8 s_n = 1$, and $(p_{n, 1}, p_{n, 2}, p_{n, 3}) \in \{1, 2\}^3$.
Note that with the chosen setup, $\theta_1$ is the most important parameter, $\theta_{2}, \theta_{3}, \theta_{4}$ are the second most important parameters etc under the prior density.

Bayesian inference is carried out for the weights $(\theta_1, \theta_2, \ldots, \theta_8)$ of the permeability field $k(x, y, z, \boldsymbol{\theta})$.
Thus, we are solving an $8$D inversion problem.
These weights follow a standard normal prior distribution, i.e., $\mu_0 = \mathrm{N}(\boldsymbol{0}, I)$.
To perform the Bayesian inference we employ both the standard and our proposed multilevel approach with the two variants, $\mathrm{MLLejaStd}$ and $\mathrm{MLLejaDV}$, considering $J = 3$.
The employed multilevel setup is outlined in \cref{tab:ml_setup_test_case_4}.
The forward model \cref{eq:test_case_4} is discretized via standard tetrahedral FE meshes $h_j$.
Moreover, the sparse grid interpolation tolerances $tol^{\mathrm{in}}_j$ and $\boldsymbol{\tau}^{\mathrm{in}}_j$ are chosen to yield quantitatively similar errors to the FE approximation for $j = 1, 2, 3$.
Finally, we choose small tolerances for quadrature to prevent the adaptive algorithm from stopping too early.
\begin{table}[tbhp]
{\footnotesize
  \caption{Multilevel setup for the 8D inversion problem with forward model \cref{eq:test_case_4}.}\label{tab:ml_setup_test_case_4}
\begin{center}
  \begin{tabular}{|c|c|c|c|c|} \hline
   Level & $h$ & $tol^{\mathrm{in}}$ & $\boldsymbol{\tau}^{\mathrm{in}}$ & $tol^{\mathrm{qu}}$ \\ \hline
    $\ell(1)$ & $h_1 = \sqrt{3}/2^{4}$ & $tol^{\mathrm{in}}_3 = 10^{-5}$ & $\boldsymbol{\tau}^{\mathrm{in}}_3 = 10^{-7} \cdot \boldsymbol{1}_8$ & $tol^{\mathrm{qu}}_3 = 10^{-9}$ \\
    $\ell(2)$ & $h_2 = \sqrt{3}/2^{5}$ & $tol^{\mathrm{in}}_2 = 10^{-4}$ & $\boldsymbol{\tau}^{\mathrm{in}}_2 = 10^{-6} \cdot \boldsymbol{1}_8$ & $tol^{\mathrm{qu}}_2 = 10^{-8}$ \\
    $\ell(3)$ & $h_3 = \sqrt{3}/2^{6}$ & $tol^{\mathrm{in}}_1 = 10^{-3}$ & $\boldsymbol{\tau}^{\mathrm{in}}_1 = 10^{-5} \cdot \boldsymbol{1}_8$ & $tol^{\mathrm{qu}}_1 = 10^{-7}$ \\ \hline
  \end{tabular}
\end{center}
}
\end{table}

The observation data consists of $729$ measurements at $\{0.1, 0.2, \ldots, 0.9\}^3 \in \Omega$, stemming from the FE solution of the forward model discretized using a finer mesh width $h = \sqrt{3}/2^7$ and assuming measurement noise $\eta \sim \mathrm{N}(\boldsymbol{0}, 0.1^2I)$.
In addition, $\boldsymbol{\theta}_{\mathrm{true}}$ is drawn from the standard multivariate Gaussian density, i.e., 
\begin{equation*}
\boldsymbol{\theta}_{\mathrm{true}} = (0.3015,  0.6578, -0.5002,  0.4608,  1.1345, 0.5447, -1.5353, -0.1689).
\end{equation*}

The QoI is again the expectation of the posterior density, $\mathbb{E}_{\pi^{\boldsymbol{y}}}[\boldsymbol{\theta}]$.
We begin with level $\ell(1)$ in the standard multilevel approach.
Both the expectation and the covariance matrix of the corresponding posterior are computed since we need these evaluations at $\ell(2)$.
We obtain, however, an indefinite covariance matrix with a negative variance for $\theta_1$.
This is mainly due to the limitations of the standard approach: adaptive sparse grid quadrature w.r.t. the prior density becomes challenging when the complexity of the underlying Bayesian inverse problem increases.
To overcome this limitation, we employ instead standard sparse grids of a priori fixed levels having sufficiently many points to guarantee a positive definite covariance matrix.
In particular, we consider an interpolation grid of level $10$ for interpolation ($24310$ grid points) and a quadrature grid of the same level comprising $598417$ points.
Since our goal is to compare multilevel methods based on adaptive sparse grid algorithms, we do not perform the standard multilevel approach on the remaining two levels using a priori chosen sparse grids, but rather focus only on the two variants of our proposed approach starting from the Gaussian approximation of the posterior at level $\ell(1)$ obtained with the aforementioned standard sparse grids.
Moreover, evaluating the FE discretizations depending on $h_2$ and $h_3$ on the standard sparse grid of level $10$ ($24310$ evaluations) requires a significant computational cost.

We first compute a reference MH solution with $10^5$ samples.
The Gaussian proposal has covariance matrix $C_{\mathrm{MH}} = 0.5 I$. 
To reduce the burn-in, the chain is started from $\boldsymbol{\theta}_{\mathrm{true}}$.
The acceptance rate is $24\%$.
Afterwards, we employ MLLejaStd and MLLejaDV as described above.
The results are showed in \cref{tab:res_test_case_4}.
The two variants of our approach produce results comparable to the reference solution, thus making our proposed approach competitive with sampling methods in this test case as well.
Observe, however, that the accuracy of all estimates deteriorates compared with $\boldsymbol{\theta}_{\mathrm{true}}$.
This is because the likelihood becomes less informative as the index increases from $1$ to $8$: the likelihood updates the prior very well for the first direction, relatively well for the next three, and almost not at all for the last four directions.
Nevertheless, since the last four directions are the least important by construction \cref{eq:diff_test_case_4} we expect that having not accurate corresponding mean estimates will not be too significant.
\begin{table}[tbhp]
{\footnotesize
  \caption{Estimation of the posterior's mean value for the fourth test case using a referece MH solution with $10^5$ samples and the two variants of our proposed multilevel approach for Bayesian inversion.} \label{tab:res_test_case_4}
\begin{center}
  \begin{tabular}{|c|c|c|c|c|c|c|c|c|} \hline
   Method & $\mathbb{E}_{\pi^{\boldsymbol{y}}}[\theta_1]$ & $\mathbb{E}_{\pi^{\boldsymbol{y}}}[\theta_2]$ & $\mathbb{E}_{\pi^{\boldsymbol{y}}}[\theta_3]$  & $\mathbb{E}_{\pi^{\boldsymbol{y}}}[\theta_4]$  & $\mathbb{E}_{\pi^{\boldsymbol{y}}}[\theta_5]$  & $\mathbb{E}_{\pi^{\boldsymbol{y}}}[\theta_6]$  & $\mathbb{E}_{\pi^{\boldsymbol{y}}}[\theta_7]$  & $\mathbb{E}_{\pi^{\boldsymbol{y}}}[\theta_8]$ \\ \hline
    MH & $0.2532$ & $0.2123$ & $-0.1363$  & $0.1326$  & $0.1486$  & $0.0753$  & $-0.1584$  & $ 0.0066$ \\
     MLStd & $0.2642$ & $0.2111$ & $-0.1630$  & $0.1539$  & $0.1429$  & $0.0670$  & $-0.1816$  & $-0.0053$ \\
      MLDV & $0.2620$ & $0.2114$ & $-0.1600$  & $0.1542$  & $0.1448$  & $0.0689$  & $-0.1808$  & $-0.0050$ \\ \hline
  \end{tabular}
\end{center}
}
\end{table}

\begin{figure}[htbp]
  \centering
  \includegraphics[width=0.75\textwidth]{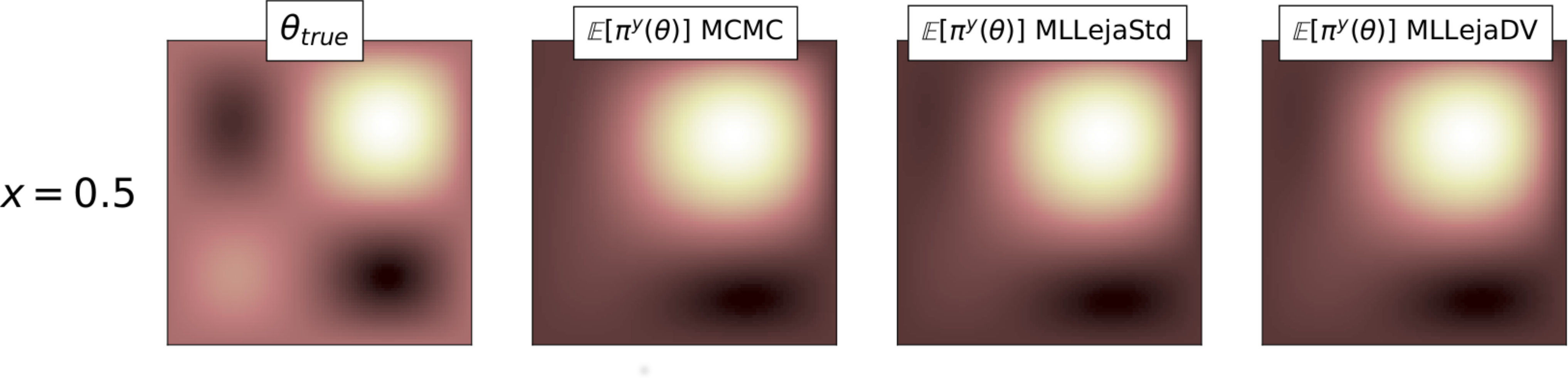}
  \includegraphics[width=0.75\textwidth]{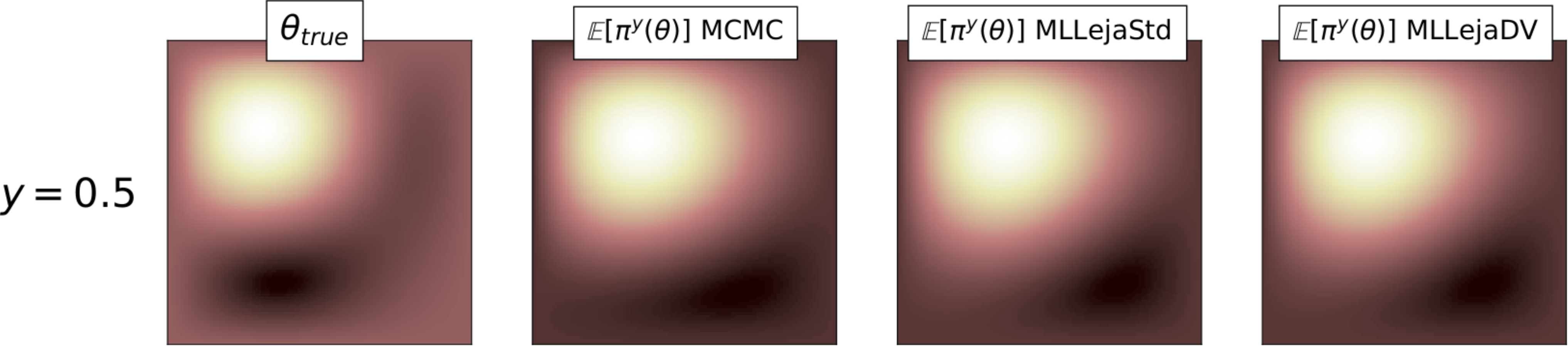}
  \includegraphics[width=0.75\textwidth]{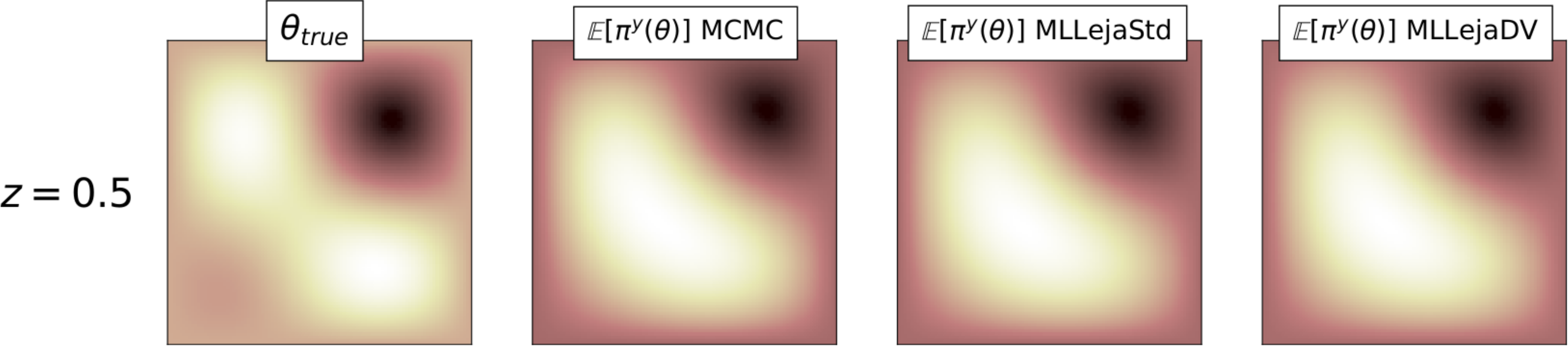}
  \caption{$2$D slices of the $3$D permeability field from \cref{eq:diff_test_case_4} parametrized using  $\boldsymbol{\theta}_{\mathrm{true}}$ and the mean estimates from \cref{tab:res_test_case_4}.
  Top: $yz$ slice taking $x = 0.5$.
  Middle: $xz$ slice when $y = 0.5$.
  Bottom: $xy$ slice fixing $z = 0.5$.}  \label{fig:rf_slices_test_case_4}
\end{figure}

To assess the quality of the expectation estimates, we use them to represent the permeability field as $k(x, y, z, \mathbb{E}_{\pi^{\boldsymbol{y}}}[\boldsymbol{\theta}])$ and we compare the results with $k(x, y, z, \boldsymbol{\theta}_{true})$.
In \cref{fig:rf_slices_test_case_4} we depict $2$D slices of the field in which the spatial coordinates $x$ (top), $y$ (middle) and $z$ (bottom) are fixed respectively to $0.5$.
We observe that having inaccurate estimates for the latter four components of $\boldsymbol{\theta}$ does not significantly affect the estimation of the true permeability field.
This is due to having good estimates for the first components of $\boldsymbol{\theta}_{\mathrm{true}}$, which are the most important by construction.
\begin{figure}[htbp]
  \centering
  \includegraphics[width=0.6\textwidth]{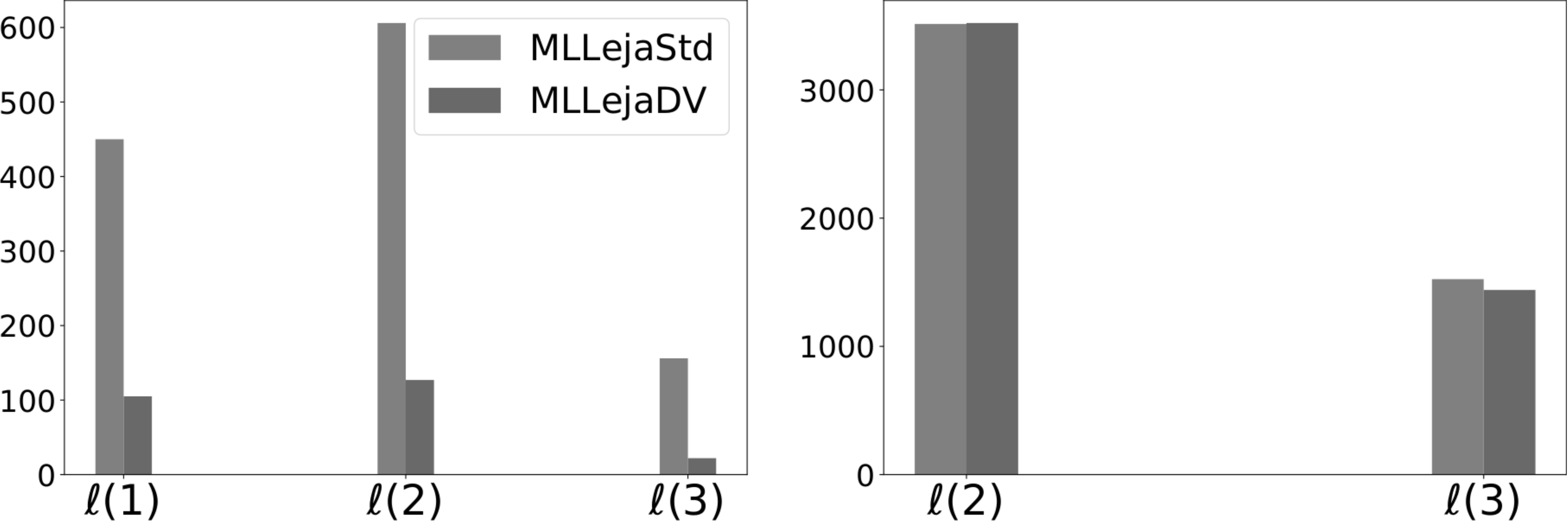}
  \caption{Left: total number of forward model evaluations needed by the two variants of our proposed approach, MLLejaStd and MLLejaDV, at all three levels in the $8$D inversion problem with forward model \cref{eq:test_case_4}. 
  			Right: total number of quadrature nodes used in all three in MLLejaStd and MLLejaDV at levels $\ell(2)$ and $\ell(3)$.}
  \label{fig:prob3D8D_cost}
\end{figure}

In \cref{fig:prob3D8D_cost} the costs for interpolation (left) and quadrature (right) are shown.
Note that for interpolation we show costs for level $\ell(1)$ as well because evaluations of the forward PDE discretized using $h_1$ are needed at $\ell(2)$.
MLLejaDV is cheaper than MLLejaStd at all three levels, requiring about $4.3$ times fewer evaluations at level $\ell(1)$, $4.8$ times fewer evaluations on level $\ell(2)$ and $7$ times fewer evaluations at level $\ell(3)$.
Observe that the overall interpolation costs are very small given that we have an $8$D inversion problem at hand.
For example, at level $\ell(3)$, MLLejaDV requires only $22$ PDE evaluations.
The maximum reached level in the corresponding multiindex set is $4$ and all its multiindices have components larger than $1$ only in the first four directions.
Indeed, the directional variances-based algorithm detects that the latter four directions are unimportant, thus invests effort only in the first four directions.
Therefore, we see once again that using the enhanced adaptive algorithm for adaptive sparse grid interpolation leads to significant cost savings.
For quadrature, the total number of evaluations of the interpolation surrogates are similar.

% conclusions
\section{Conclusions}\label{sec:conc}
We proposed a novel multilevel Leja algorithm for computing posterior approximations in computationally expensive, higher-dimensional Bayesian inverse problems.
At each level adaptive sparse grid interpolation is employed to find a surrogate of the potential function, and adaptive sparse grid quadrature is then used to perform all integration operations with respect to the posterior.
We considered two adaptive strategies for interpolation: (i) a standard method and (ii) an enhanced adaptive algorithm in which directional variances are used to ensure that only the most important stochastic directions are refined.
The backbone of the proposed approach is the sequential update of the prior density.
In this way, we can create weighted (L)-Leja points in areas of high posterior probability.
Numerical experiments with elliptic inverse problems in 2D and 3D space show that the sequential update of the prior leads to considerably fewer model evaluations compared to the standard multilevel approach which employs the prior density at all levels.
%For example, in the second test case, the source inversion problem with one source in a two dimensional spatial domain, we obtained up to $20$ times fewer interpolation points and up to $12.5$ fewer quadrature nodes compared to the standard multilevel approach.
%Our results thus show that the proposed multilevel method is suitable for computationally expensive, higher-dimensional Bayesian inverse problems.
We remark that the proposed approach is not designed to handle well multimodal posterior densities.
In future research we will extend it such that it employs more general nonlinear mappings, e.g., transport maps, to accurately approximate arbitrary, multimodal posterior densities.

% acknowledgements
\section*{Acknowledgements}
IGF thankfully acknowledges the support of the German Academic Exchange Service (DAAD).  
IGF, JL, and EU would like to thank the Isaac Newton Institute for Mathematical Sciences for support and hospitality during the programme \emph{Uncertainty quantification for complex systems: theory and methodologies} when work on this paper was undertaken. 

% bibliography
% \clearpage
\bibliographystyle{siamplain}
\bibliography{ms}

\end{document}